\documentclass[preprint2]{aastex61}

\usepackage{graphicx}
\usepackage[space]{grffile}
\usepackage{latexsym}
\usepackage{textcomp}
\usepackage{longtable}
\usepackage{multirow,booktabs}
\usepackage{amsfonts,amsmath,amssymb}
\usepackage{natbib}
\usepackage{url}
\usepackage{hyperref}
\hypersetup{colorlinks=false,pdfborder={0 0 0}}
\newif\iflatexml\latexmlfalse
\DeclareGraphicsExtensions{.pdf,.PDF,.png,.PNG,.jpg,.JPG,.jpeg,.JPEG}

\usepackage[utf8]{inputenc}
\usepackage[ngerman,english]{babel}

\newcommand\aastex{AAS\TeX}

\received{05/12/2017}
\revised{10/05/2017}
\accepted{10/21/2017}
\submitjournal{ApJ}

\shorttitle{\aastex\ Toward the analysis of JWST exoplanet spectra}
\shortauthors{Baudino et al.}

\begin{document}

\title{Toward the analysis of JWST exoplanet spectra:\\
       Identifying troublesome model parameters}

\correspondingauthor{Jean-Loup Baudino}
\email{jean-loup.baudino@physics.ox.ac.uk, jean-loup.baudino@obspm.fr}

\author[0000-0003-4061-2514]{Jean-Loup Baudino}
\affil{Irfu, CEA, Universit\'e Paris-Saclay, 91191 Gif-sur Yvette, France}
\affil{AIM, Universit\'e Paris Diderot, 91191 Gif-sur-Yvette, France}
\affil{Department of Physics, University of Oxford, Oxford, UK}

\author{Paul Molli\`ere}
\affil{Max Planck Institute for Astronomy, Königstuhl 17, D-69117 Heidelberg, Germany}

\author{Olivia Venot}
\affil{Laboratoire Interuniversitaire des Syst\`emes Atmosph\'eriques (LISA), CNRS UMR 7583, Universit\'e Paris-Est Cr\'eteil, Universit\'e Paris Diderot, 61 avenue du G\'en\'eral de Gaulle, 94010 Cr\'eteil Cedex, France}
\affil{Instituut voor Sterrenkunde, Katholieke Universiteit Leuven, Leuven, Belgium}

\author[0000-0001-6172-3403]{Pascal Tremblin}
\affil{Astrophysics Group, University of Exeter, Exeter, EX4 4QL, UK}
\affil{Maison de la Simulation, CEA, CNRS, Univ. Paris-Sud, UVSQ, Universit\'e Paris-Saclay, 91191 Gif-sur-Yvette, France}

\author[0000-0002-5433-5661]{Bruno B\'ezard}
\affil{LESIA, Observatoire de Paris, PSL Research University, CNRS, Sorbonne Universit\'es, UPMC Univ. Paris 06, Univ. Paris Diderot, Sorbonne Paris Cit\'e, 5 place Jules Janssen, 92195 Meudon, France}

\author{Pierre-Olivier Lagage}
\affil{Irfu, CEA, Universit\'e Paris-Saclay, 91191 Gif-sur Yvette, France}
\affil{AIM, Universit\'e Paris Diderot, 91191 Gif-sur-Yvette, France}



\begin{abstract}

Given the forthcoming launch of the James Webb Space Telescope 
(JWST) which will allow observing exoplanet atmospheres with 
unprecedented signal-over-noise ratio, spectral coverage and 
spatial resolution, the uncertainties in the atmosphere 
modelling used to interpret the data need to be assessed. 
As the first step, we compare three independent 
1D radiative-convective models: \emph{ATMO}, \emph{Exo-REM} and 
\emph{petitCODE}. We identify differences in physical and chemical 
processes taken into account thanks to a benchmark protocol we 
developed. We study the impact of these differences on the 
analysis of observable spectra. We show the importance of 
selecting carefully relevant molecular linelists to compute the 
atmospheric opacity. Indeed, differences between spectra 
calculated with Hitran and ExoMol exceed the expected 
uncertainties of future JWST observations. We also show the 
limitation in the precision of the models due to uncertainties 
on alkali and molecule lineshape, which induce spectral 
effects also larger than the expected JWST uncertainties. We 
compare two chemical models, \emph{Exo-REM} and 
\emph{Venot Chemical Code}, which do not lead to significant 
differences in the emission or transmission spectra. We discuss 
the observational consequences of using equilibrium or out-of-
equilibrium chemistry and the major impact of phosphine, 
detectable with the JWST.Each of the models has benefited from 
the benchmarking activity and has been updated. The protocol 
developed in this paper and the online results can constitute a 
test case for other models.

\end{abstract}




\section{Introduction}

Since the publication of the detection of an extrasolar planet 
orbiting the 51 Peg star by \cite{Mayor1995b}, the field of 
exoplanet detection has developed very rapidly; more than 3700 
exoplanets have been detected so far (e.g. \url{exoplanet.eu}). 
The field is now shifting more and more from the detection of 
exoplanets to the characterization of known exoplanets, especially 
their atmosphere, thanks to spectroscopic observations in the 
visible and infrared. One of the prime facilities used so far to 
observe the atmosphere of transiting exoplanets has been the 
Hubble Space Telescope (HST) \citep[see the review by][ and 
references therein]{Deming2017}. Thanks to the use of the 
so-called scanning observing techniques with the WFC3 instrument, 
as well as sophisticated data reduction methods, variation of the 
star flux due to the transit of an exoplanet has been detected 
down to a few tens of ppm \citep{Kreidberg2014, Tsiaras2016}. In 
the near future, the James Webb Space Telescope (JWST, see 
\url{jwst.stsci.edu}), thanks to a large collecting area and a 
suite of state to the art instruments covering a large wavelength 
range (0.6-28 microns), will be a key machine to study the 
atmosphere of exoplanets \citep[e.g. ][]
{Greene2016,Molli_re_2017}. The JWST will not only be used 
to study transiting/eclipsing exoplanets, but also exoplanets 
observed by direct imaging, thanks to its high angular 
resolution. For those exoplanets, far enough from the star to 
allow spectroscopic observations, spectra of exoplanet emission 
with a signal over noise ratio greater than 100 are expected.  

Atmospheric models are needed to interpret the observations 
quantitatively. Several models have been developed over the years 
\citep[for example see the reviews by ][and the references 
there-in]{Helling2008h, Marley2015, Hubeny2017}. Given the high 
observational precision achieved nowadays with current facilities 
and in the near future with the JWST, the question of comparing 
the uncertainties in the model predictions with the precision of 
the observations has to be considered. To tackle the question, we 
have, as a first step, only considered one dimensional models. To 
investigate the uncertainties in the models prediction, our 
approach has been to start with the comparison of the results from 
three models developed independently: the \emph{petitCODE} model 
\citep{Molli_re_2015,Molli_re_2017}, the \emph{ATMO} model 
\citep{Tremblin_2015} and the \emph{Exo-REM} model 
\citep{Baudino_2015}. To identify the differences between 
the models and their impact on the predicted spectra, we have 
adopted a benchmark approach to give a common basis to the models 
allowing disentangling the impact of such or such differences.

In the second Section of the paper, the current version of the 
codes are briefly discussed and their differences outlined. The 
next section describes the protocol developed to benchmark the 
models. The results of the benchmark are discussed in 
Sect.~\ref{benchmark}. In Sect.~\ref{keysSect}, we focus on the 
evaluation of the uncertainties due to the ways the shape of 
resonant line far wings of alkalies are treated and due to the 
molecule far wing lineshape in use. In Sect.~\ref{nonequilibrium}, 
we consider out-of-equilibrium chemistry 
\citep[see ][]{Moses2011, Venot_2012, Line2013, Zahnle_2014, Hu2015} 
using two approaches coming from \emph{Exo-REM} and 
\cite{Venot_2012} and the observational consequences of this 
phenomenon. Then, in Sect.~\ref{detectability}, we compare 
the deviation due to the different modelling approaches to the 
expected JWST uncertainties. The conclusions and perspectives 
are drawn in Sect.~\ref{discuss}.

\section{Description of the radiative-convective equilibrium models \label{modelsDesc}}
Before focusing on the benchmark itself, we describe here the 
current state of the three radiative-convective equilibrium models 
which have been benchmarked. This is the state of the models 
taking into account the evolution resulting from the benchmark. 

These models compute the 1D vertical structure of the atmosphere 
of giant planets and generate emission and/or transmission spectra 
given a set of input parameters: the effective temperature 
$T_\mathrm{eff}$, the surface gravity $\log_{10}(g)$, the radius 
and the elemental composition.\\

The atmosphere is discretised in a number of layers (50, 64 and 
120 respectively for \emph{ATMO}, \emph{Exo-REM} and 
\emph{petitCODE}). The models calculate the net energy flux as a 
function of pressure level using the radiative transfer equation, 
and solve it iteratively for radiative-convective equilibrium, 
i.e. conservation of the flux. The codes also incorporate a 
thermochemical model that calculates layer-by-layer the molecular 
mole fractions, given the elemental abundances and temperature 
profile. Heating sources, which can be internal (e.g. coming from 
the formation and evolution process) or external (e.g. coming from 
the star), are taken into account. 

All the models take the internal heating due to the contraction of 
the planet into account:
\begin{itemize}
   \item \emph{ATMO} and \emph{petitCODE} consider additional 
   heating by the star radiation and are thus adapted to model 
   close-in exoplanets and to calculate transit spectra.
   \item \emph{Exo-REM} is only suited to young giant exoplanets 
   for which stellar heating can be neglected with respect to the 
   internal heat flux. This model is not able to compute 
   transmission spectra.
\end{itemize}

One of the key drivers to determine the structure of an atmosphere 
is its opacity, which depends on its atomic and molecular 
composition, which, in turn, depends on the chemistry at work in 
the atmosphere. 

\subsection{Opacities}

All the models consider continuum collisional induced absorption 
(CIA) for H$_2$--H$_2$ and H$_2$--He from the same references, 
atomic absorption coming from Na and K with various approaches for 
the resonant lines (see Sect.~\ref{alkali}), molecular absorption 
coming from H$_2$O, CH$_4$, CO, CO$_2$, NH$_3$, PH$_3$, TiO, and 
VO. A lot of work has been recently devoted to update molecular 
linelists, especially by the Exomol team (see \url{exomol.com}). 
The references for the molecular linelists used in the various 
models are given in Table~\ref{difopaclist}. It is interesting to 
note that, except for CO$_2$, at least two of the three models use 
the same linelist for a given molecule, which makes possible to 
infer the influence of using different linelists.  Absorption 
coefficients are computed for a broadening due to:

\begin{itemize}
   \item terrestrial atmosphere (air) for \emph{petitCODE}
   \item a mix of H$_2$ and He for \emph{ATMO}, \emph{Exo-REM}.
\end{itemize}

\begin{table*} 
\caption{References of the different sources of molecular opacities 
for the three models considered here (\emph{ATMO}, \emph{Exo-REM}, 
\emph{petitCODE}) \label{difopaclist} }
    \begin{tabular}{ c | c | c | c }
         \hline\hline
         Opacities & \multicolumn{3}{|c}{ Common references }\\ 
         \hline 
         CH$_4$  & \multicolumn{3}{|c}{ \cite{Yurchenko_2014}}\\ 
         \hline
         {TiO and VO} & \multicolumn{3}{|c}{ \cite{Plez_1998} (with update from private communication)} \\
         \hline
         H$_2$--H$_2$, & \multicolumn{3}{|c}{ HITRAN \citep{Richard_2012}}\\
         H$_2$--He &  \multicolumn{3}{|c}{ \citet{Borysow_2001} and \citet{Borysow_2002}}\\ 
         \hline
         Opacities & \emph{ATMO} & \emph{Exo-REM} & \emph{petitCODE} \\ 
         \hline 
         {H$_2$O} & {\cite{Barber_2006}} & { \cite{Rothman_2010}} &  { \cite{Rothman_2010}} \\\hline 
         {CO}  & { \cite{Rothman_2010}} & { \cite{Rothman_2010}} & { \cite{Rothman_2010},\cite{Kurucz1993}} \\\hline
         {CO$_2$} & { \cite{Tashkun_2011}} & { \cite{Rothman_2013} } & { \cite{Rothman_2010}}\\\hline
         {NH$_3$} & {\cite{Yurchenko_2011} } & { \cite{Yurchenko_2011} } & { \cite{Rothman_2013}}\\\hline
         {PH$_3$} & { \cite{Sousa_Silva_2014} } & { \cite{Sousa_Silva_2014} } &  { \cite{Rothman_2013}}\\\hline  
    \end{tabular} 
    
\end{table*}

Note that PH$_3$ and CO$_2$ have been added in \emph{Exo-REM} 
since the first version described in \cite{Baudino_2015}.

After considering the linelists sources we focus on how each 
model computes the shape of the wings of each line (except for 
alkalies fully described in Section~\ref{alkali} because the 
different approaches for the wings treatment of the atomic lines 
used in the models induce significant differences in the model 
predictions). Our three codes use Voigt profile but with various 
"cut-off" (or sub-Lorentzian lineshape) implementations:

\begin{itemize}
   \item For \emph{ATMO}, the line cut-off is described in detail
    in \cite{Amundsen_2014}. The cut-off distance is calculated 
    on-the-fly by estimating when the line mass absorption 
    coefficient has reached a critical value.
   \item For \emph{petitCODE} and \emph{Exo-REM} the applied 
   sub-Lorenzian lineshape comes from \cite{Hartmann_2002} for all 
   molecules except for CO$_2$ for which a profile from 
   \cite{Burch_1969a} is used \citep[see ][]{Molli_re_2017}.
\end{itemize}

For fast computation, the opacity is treated by 
using opacity distribution functions and making use of the 
correlated-k assumption. This technique is very useful, especially 
to manage the large number of lines in modern molecular 
linelists. In our case, all models use correlated-k 
coefficients computed molecule-by-molecule. 

Our three models combine the molecules assuming no correlation 
between species in any spectral interval.  This is done using the 
method named "reblocking of the joint k--distribution" by 
\cite{Lacis_1991} and more extensively described in 
\citet{Molli_re_2015} ("R1000" method in their Appendix B.2.1) 
with a few differences for \emph{ATMO} as described in  
\cite{Amundsen_2014, Amundsen_2016}.\\

\subsection{Equilibrium chemistry}

The three models compute the equilibrium chemistry using various 
methods.

\emph{ATMO} and \emph{petitCODE} use a Gibbs energy minimisation 
scheme following the method of \cite{gordon_1994}.

\emph{ATMO} uses the same thermochemical data as \cite{Venot_2012} 
in the form of NASA polynomial coefficients \citep[see ][]
{McBride_1993}. The Gibbs minimizing method allows for depletion 
of gas phase species due to condensation and is included in the 
chemical solver.

\emph{petitCODE} uses the same NASA polynomials, but extends them 
to the low temperatures ($<$200~K) using the JANAF database 
\citep[see also Appendix A2 of ][]{Molli_re_2017}.

\emph{Exo-REM} solves the thermochemical equilibrium 
layer-by-layer using on-line JANAF data from \cite{chase_1998}. 
\emph{Exo-REM} does not use a general Gibbs energy minimisation 
scheme, but a simplified approach. The molecules are  grouped by 
independent sub-sets \citep[see Tab.~2 in][]{Baudino_2015}. In a 
given sub-set, the molecular mole fractions are found by solving a 
system of equations consisting of equilibrium constants and 
element conservation.
When condensation occurs in a layer, the elements trapped in the 
condensed species are considered as lost for the chemistry 
occuring at higher layers. Doing so, ExoREM includes the 
so-called cold trap phenomenon in contrast to \emph{ATMO} and 
\emph{petitCODE}.

\section{Benchmark protocol \label{benchmarkcond}}

The two following sections 
(Sect.~\ref{benchmarkcond}~and~\ref{benchmark}) are focused on the 
benchmark itself. First, we define a "common" model to compare the 
results of our models (Sect.~\ref{mincore}) and the test protocol 
used to obtain the results 
(Sect.~\ref{sameptcond}~and~\ref{targetcond}). Then we compare the 
results of this benchmark in Sect.~\ref{benchmark}.

\subsection{Definition of a "common" model \label{mincore}}

The starting point of a benchmark is the definition of a common 
set of parameters and of physical, chemical processes to be taken 
into account. Given that one of the key parameters determining the 
vertical structure of an atmosphere is its atmosphere composition, 
we have first to define a common set of absorbers.\\

The common model considers the opacities from the most important 
molecules and atoms: NH$_3$, CH$_4$, CO, CO$_2$, H$_2$O, PH$_3$, 
Na, K, as well as the collision-induced absorption for 
H$_2$--H$_2$ and H$_2$--He. For the alkalies we use the following 
H$_2$ broadening coefficients:   

\begin{itemize}  
   \item Na: $0.37 \times (T/296)^{-0.65} $cm$^{-1} $atm$^{-1}$ fitted to \cite{Allard_2012},  
   \item K: $0.40 \times (T/296)^{-0.64} $cm$^{-1} $atm$^{-1}$ fitted to\cite{Allard_2016}.  
\end{itemize}

For the far wings we use a Voigt profile up to 4500 cm$^{-1}$ from 
line center for all Na and K lines and zero absorption beyond. 
Other line shapes used in the literature are explored in 
sect.~\ref{alkali}.TiO and VO are not taken into-account to keep a 
minimal "common" model; we consider them only in the case of an 
extremely hot atmosphere. We have not considered the presence of 
clouds. \\   

The abundance of the various absorbers depends on the chemical 
reactions at work in the atmosphere; we have considered a 
chemistry including the following reactant species: H$_2$, H, He, 
H$_2$O, H$_2$O(s), CO, CO$_2$, CH$_4$, NH$_3$, N$_2$, Na$_2$S(s), 
H$_2$S, Na, HCl, NaCl, K, KCl, KCl(s), NH$_4$Cl(s), SiO(s), Mg, 
Mg$_2$SiO$_4$(s), MgSiO$_3$(s), SiO$_2$(s), Fe, Fe(l), PH$_3$, 
H$_3$PO$_4$(l), P, P$_2$, PH$_2$, CH$_3$, SiH$_4$, PO.

Unless specified otherwise, we consider an atmosphere with a solar 
metallicity. We use solar elemental abundances from 
\cite{Asplund_2009}.\\  

By default, the radius, at 10~bars, of the exoplanet is fixed at 
1.25~$R_\mathrm{jup}$ (where $R_\mathrm{jup}$ is the 
radius of Jupiter; we use 69911~km for 1~$R_\mathrm{jup}$) and the 
distance between the exoplanet and the observer is fixed at 10~pc.

After adapting (upgrading or downgrading) our models to this 
common model, we generated spectra and atmospheric structures 
(composition and temperature) in the following conditions.

\subsection{Comparison with fixed temperature profiles \label{sameptcond}}

First, we compute the abundance profiles at chemical equilibrium 
and spectra for five given temperature profiles. Indeed our models 
are such that it is possible to impose a pressure--temperature 
profile (PT profile), and then run them without reaching the 
radiative-convective equilibrium, by just computing the chemistry 
and doing the radiative transfer only. We compare the result 
without iteration of the models in exactly the same conditions (in 
Sect.~\ref{targetcond} we apply the iterations needed for the 
full convergence of a model to radiative-convective equilibrium).

\begin{figure}[ht]
\begin{center}
\includegraphics[width=1.00\columnwidth]{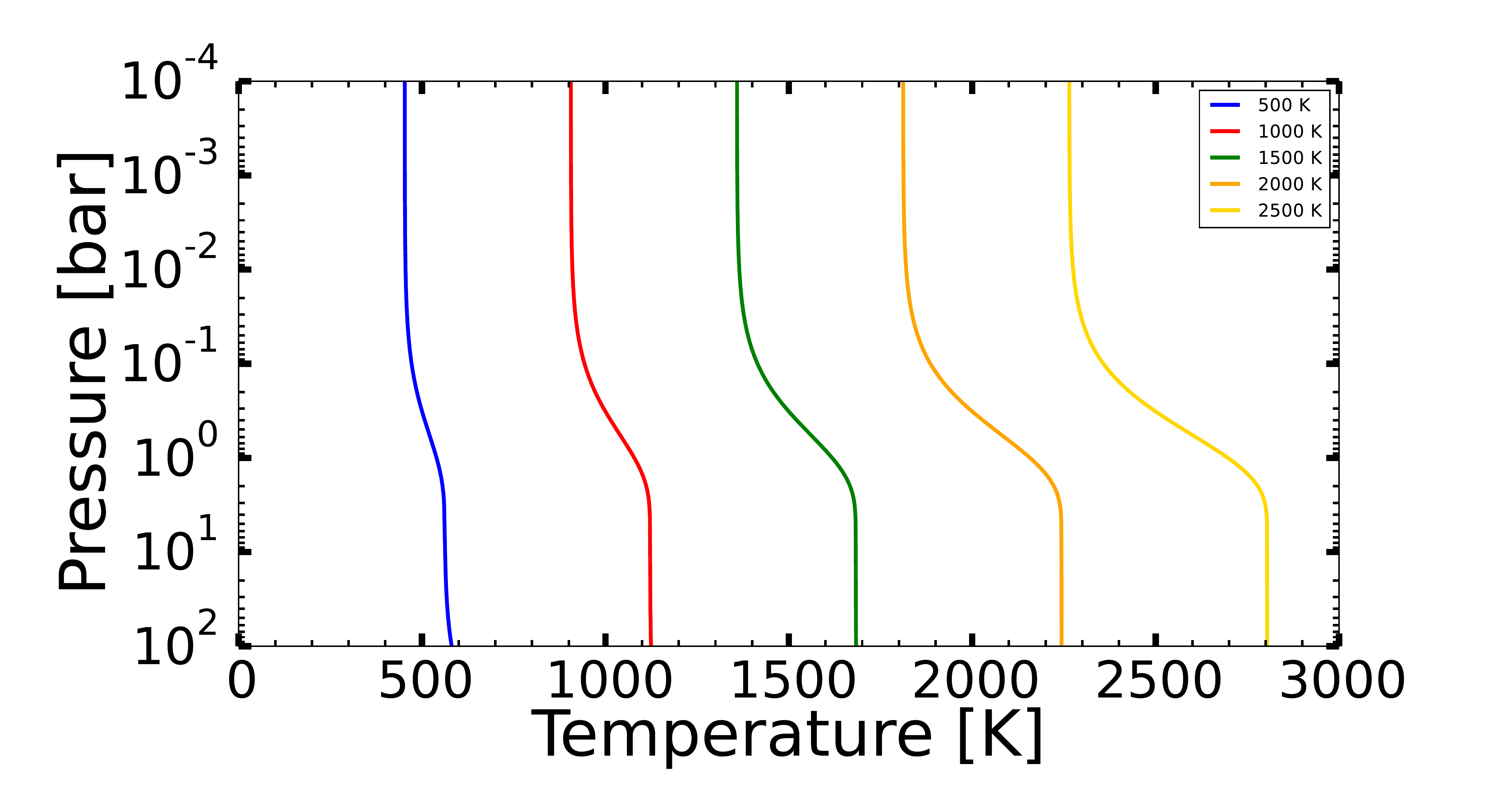}
\caption{{Input pressure-temperature profiles used in the benchmark \label{PTGuillot}}}
\end{center}
\end{figure}

We use the thermal profiles of \cite{Guillot_2010} 
(Fig.~\ref{PTGuillot}), with surface gravity 
$\log_{10}(g[\mathrm{cgs}])$=3.7, solar metallicites and effective 
temperatures (corresponding to the model of Guillot, these 
T$_\mathrm{eff}$ are not necessary consistent with models used 
here) of: 500~K, 1000~K, 1500~K, 2000~K, and 2500~K. More 
specifically, we use Equation (29) of \cite{Guillot_2010}, with 
f = 1/4, $\gamma = 0.4$, $\kappa=0.01$ and $T_{\rm int}=200$~K. 
For T$_\mathrm{eff}$ = 1000~K, we use two additional values of the 
metallicity: 3 and 30 times the solar value. 

This allows us to explore how our results compare for a broad 
range of temperatures, as well as different enrichment values.

\subsection{Comparison of self-consistent calculations \label{targetcond}}

As a next step we compare self-consistent atmospheric solutions of 
our codes, calculating radiative-convective equilibrium structures 
for known exoplanets. In the previous section, we fixed the 
same temperature-profile in each model, while in this 
section, we impose the $T_\mathrm{eff}$ itself. It means that we 
have to iterate on the P-T profile to get a self-consistent 
solution to the transfer equation. We do this for two 
direct imaging and two transiting exoplanets. We run our models, 
considering published physical parameters for the exoplanets 
(based on \url{http://exoplanet.eu}).\\  

For the two self-luminous planets we stay in the 
benchmark condition, except for the metallicity. GJ~504~b 
\citep{Kuzuhara_2013} is a cold planet with 
$T_\mathrm{eff}$=510~K, $\log(g)$=3.9 and a metallicity 
z=0.28~dex. VHS~1256--1257~b \citep{Gauza_2015} is an object 
hotter than GJ~504~b with $T_\mathrm{eff}$=880~K, $\log(g)$=4.24 
and a metallicity z=0.21~dex. In the absence of strong constraints 
on the radius, we keep the benchmark value of 
1.25~$R_\mathrm{Jup}$. \\   

For the two irradiated transiting exoplanets we keep the solar 
metallicity but adapt the radius to the published values. For  
GJ~436~b \citep{Butler_2004}, the host star parameters are: an 
effective temperature $T_\mathrm{eff}$ of 3684~K and a radius of 
0.464~solar. The planet has a $T_\mathrm{eff}$=712~K, a radius of 
0.38~$R_\mathrm{Jup}$ and a mass of 0.07~$M_\mathrm{Jup}$. The 
incidence angle of the irradiation $\cos(\theta)$=0.5.
The irradiation of the second transiting exoplanet considered 
here, WASP~12~b \citep{Hebb_2009}, is more extreme, and could 
possibly lead to a temperature inversion at high altitude. Thus 
for this planet, we decided to add the chemistry and opacities for 
TiO and VO. The host star parameters in this case are: an 
effective temperature $T_\mathrm{eff}$ of 6300~K and a radius of 
1.599~solar. The planet has an high $T_\mathrm{eff}$=2536~K, a 
radius of 1.736~$R_\mathrm{Jup}$ and a mass of 
1.04~$M_\mathrm{Jup}$. We assume an incidence angle for the 
irradiation of $\cos(\theta)$=0.5. For the Ti and V chemistry we 
considered Ti, TiO, V and VO.

\section{Results \label{benchmark}}

\subsection{Results comparison \label{resultscomp}}

In this section, we present the results of the comparison between 
our models following the benchmark protocol. We show here a 
selection of plots to illustrate these results. In 
Appendix~\ref{allfigures}, we attached the complete set of 
spectra and profiles for all the various test cases. All the 
data are available online (as supplementary material).\\

For each model, we compare the spectra, the abundance profiles of 
the absorbers and when relevant, i.e. for the four targets 
modelled in a self-consistent way, the temperature profiles. All 
spectra are plotted at the same spectral resolution (corresponding 
to that of \emph{Exo-REM}, i.e. with a step and resolution of 
20~cm$^{-1}$). The spectra are shown from near to 
mid-infrared, the wavelength range of the JWST.

\begin{figure*}[ht]
\begin{center}
\includegraphics[width=1.00\textwidth]{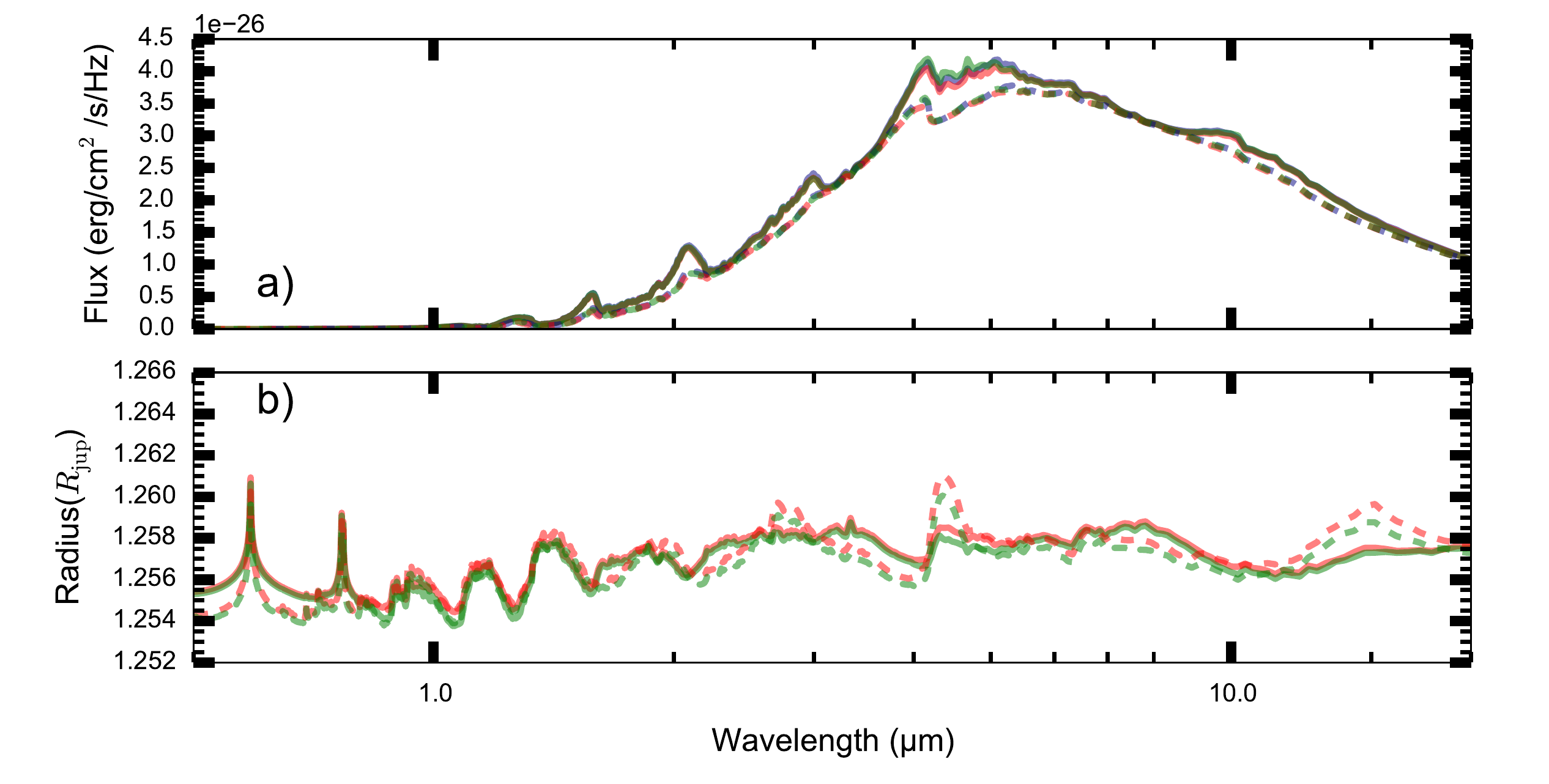}
\caption{{Emission (a) and transmission (b) spectra for models at  1000~K with solar metallicity (solid), and 1000~K with $30\times solar$  metallicity (dashed) \label{ColdTrans}}}
\end{center}
\end{figure*}

\begin{figure}[ht]
\begin{center}
\includegraphics[width=1.00\columnwidth]{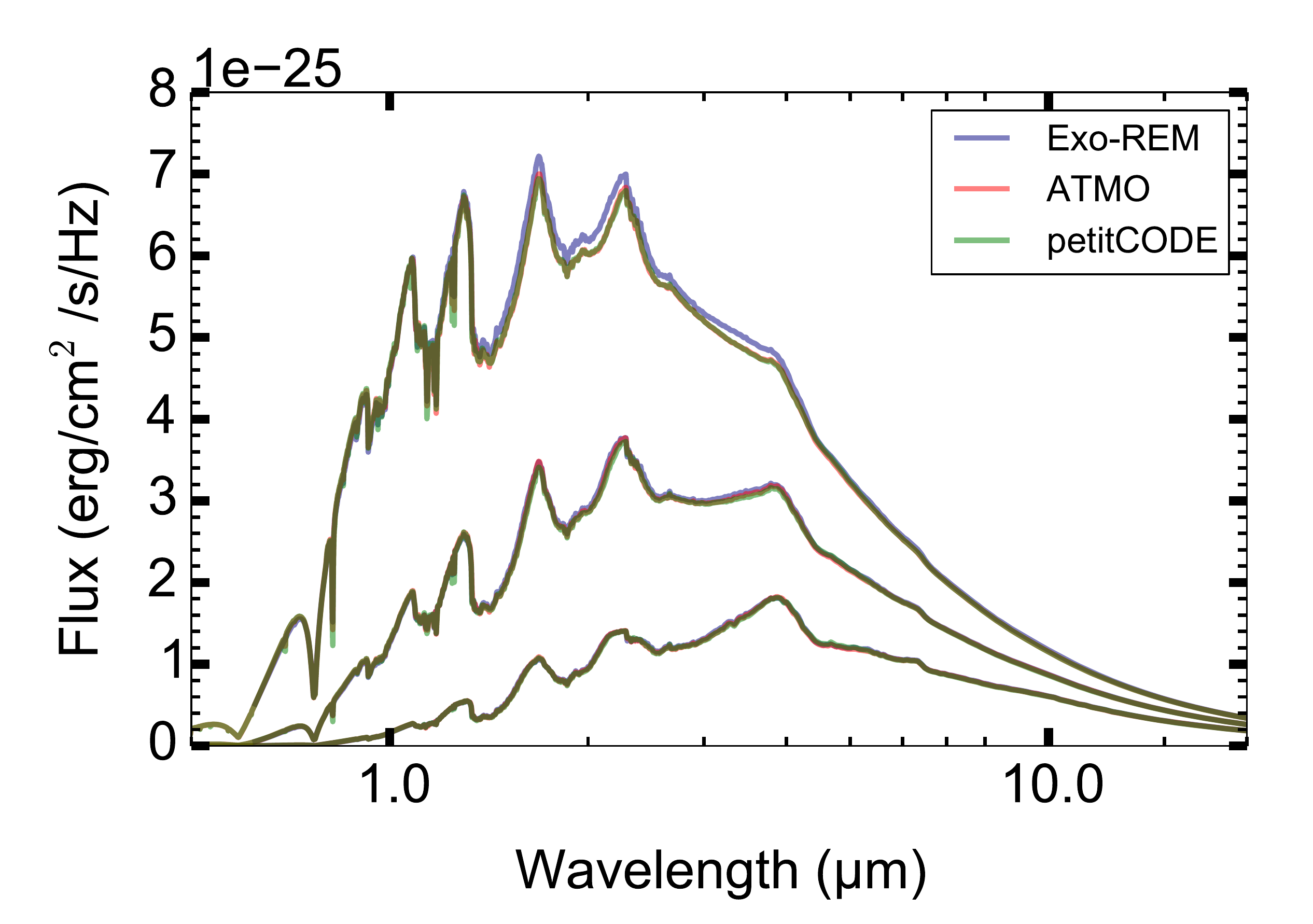}
\caption{{Emission spectra (bottom to top) for effective temperatures of 1500, 2000, 2500~K \label{HotEmis}}}
\end{center}
\end{figure}

\begin{figure*}[ht]
\begin{center}
\includegraphics[width=1.00\textwidth]{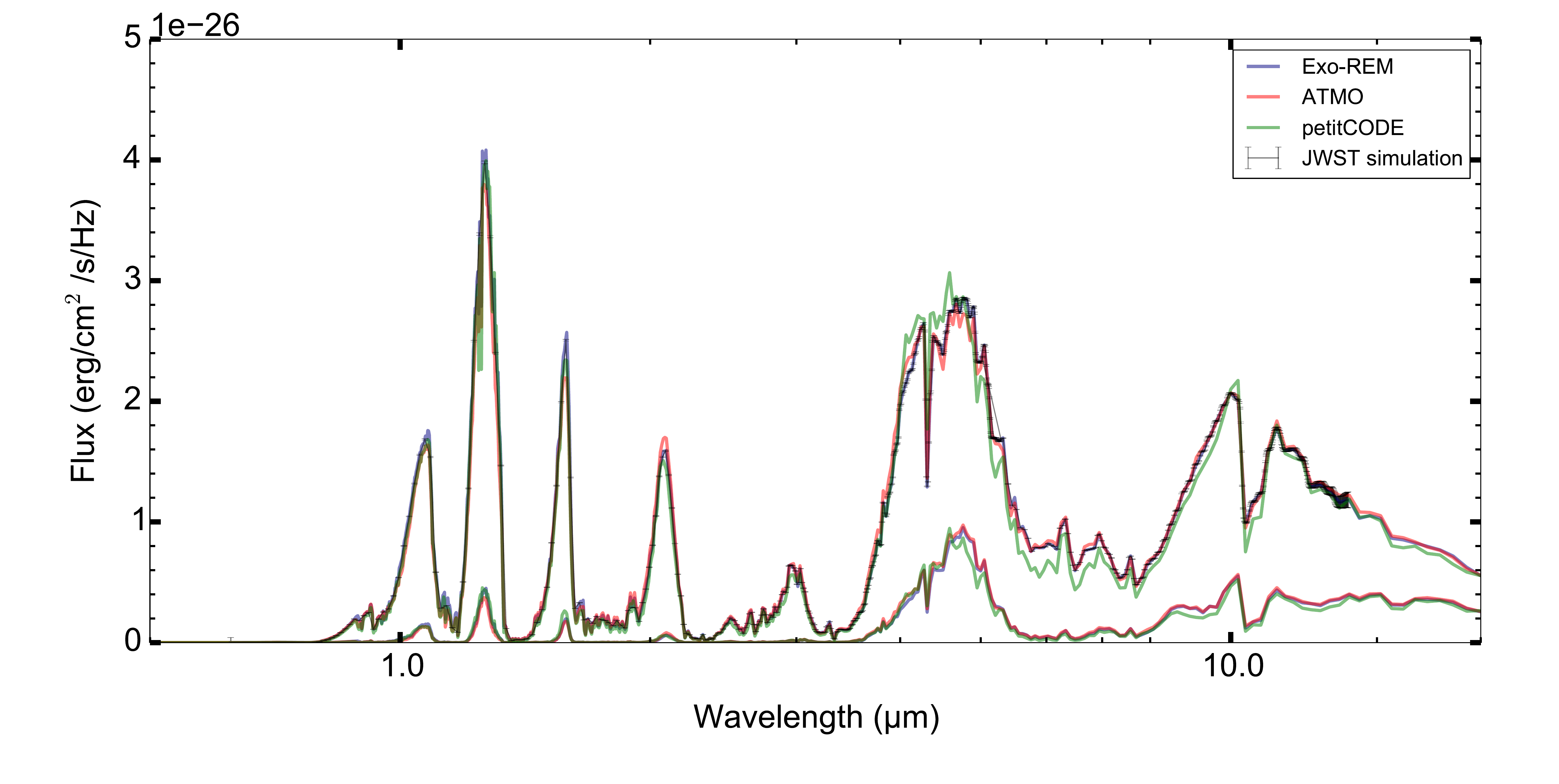}
\caption{{Emission spectra of VHS~1256-1257~b (top) and GJ~504~b (bottom). \label{EmissTest} The black uncertainties correspond to the combination of simulated NIRSpec/Prism and MIRI/LRS noise level for VHS~1256-1257~b for 0.5 hour of integration.}}
\end{center}
\end{figure*}

\begin{figure*}[ht]
\begin{center}
\includegraphics[width=1.00\textwidth]{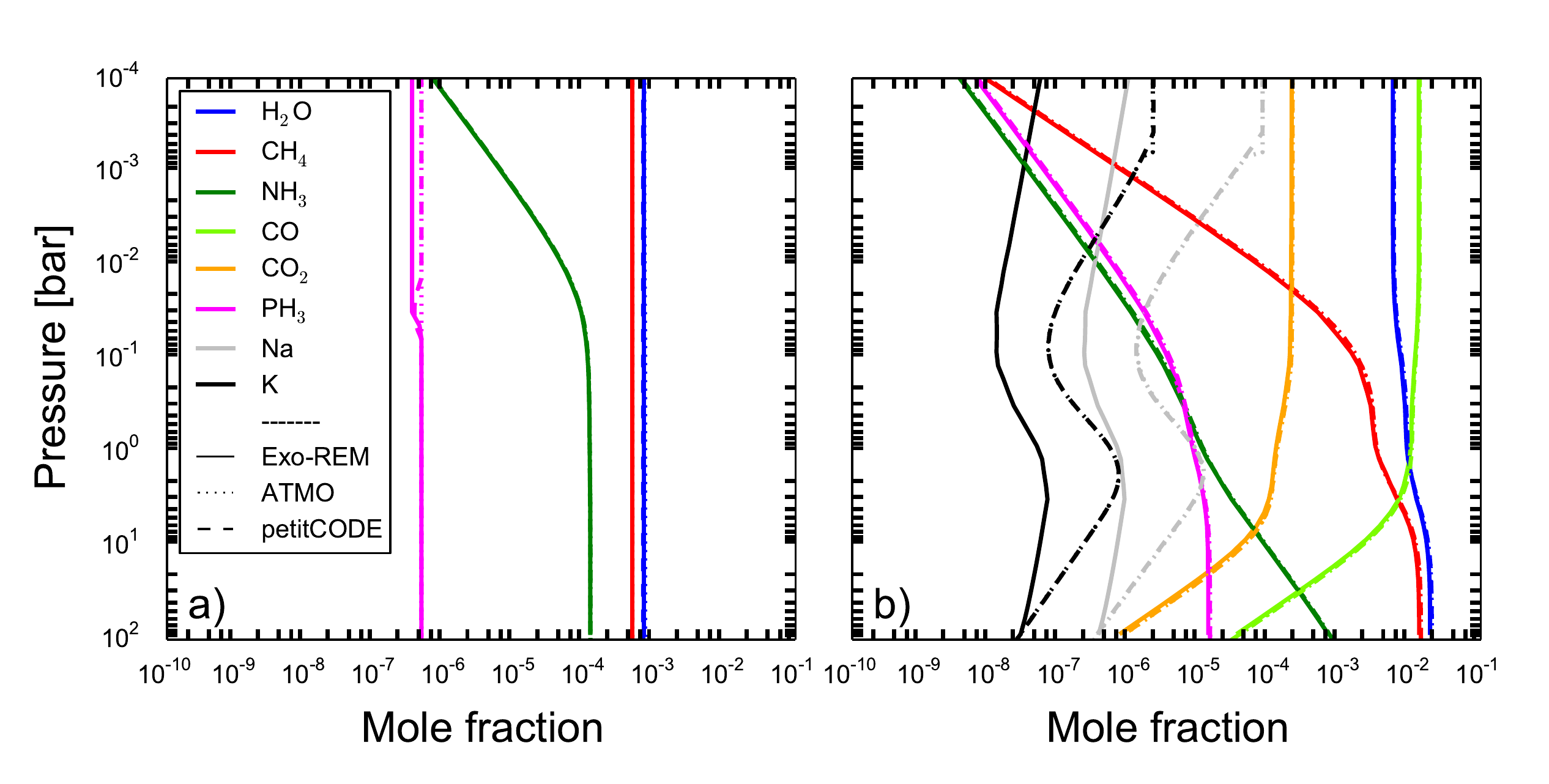}
\caption{{Abundance profiles of the defined molecules for the case with a T$_\mathrm{eff}$ =500~K at solar metallicity (a) and 1000~K at 30 $\times$ solar metallicity (b). The curves for the models are often superposed, except for alkalies where \emph{Exo-REM} is not.\label{1000KProf}}}
\end{center}
\end{figure*}

At the end of our convergence process (fully described in 
Section~\ref{stepbystep}), we reached a good agreement between all 
models in the common model conditions.
This is observable in Fig.~\ref{ColdTrans}, which shows the 
spectra of the cases with a T$_\mathrm{eff}$ = 1000~K at solar and 
super-solar metallicities. Figure \ref{HotEmis} shows the spectra 
for the three hottest cases (T$_\mathrm{eff}$ = 1500~K, 2000~K, 
2500~K) with also good agreement between \emph{ATMO}, 
\emph{petitCODE} and \emph{Exo-REM}. If we consider the general 
trend, the emission, transmission spectra and molecular 
abundances calculated by the three radiative-convective 
equilibrium models are very similar. It is especially interesting 
to see how similar spectra of exoplanet targets are without 
pre-defined temperature profile in the models 
(Fig.~\ref{EmissTest}, and in Appendix~\ref{allfigures} 
Fig.~\ref{targetsSpec}).\\

Nonetheless, we still observe some differences and we will be 
focusing on these in the following paragraphs. 

To highlight the impact of the remaining differences, we 
investigate the predicted error bars from JWST observations for 
VHS~1256--1257~b in Fig.~\ref{EmissTest}. 

To obtain these error bars, we use the JWST Estimator Time 
Calculator (\url{https://jwst.etc.stsci.edu/}) to simulate a half 
an hour observation with the NIRSpec JWST instrument using the 
prism mode and with the MIRI JWST instruments using the Low 
Resolution Spectroscopic mode. 

While \emph{ATMO} and \emph{Exo-REM} codes provide similar results 
within the error bars, \emph{petitCODE} provides results  
significantly different. In the case of Fig.~\ref{EmissTest} its 
comes mainly from linelists differences as present in the next 
section.

\subsubsection{Differences due to opacities}

A large part of the differences originates from the absorption of 
PH$_3$, more observable for low temperature cases (4-5~$\mu$m, 
Fig.~\ref{EmissTest} and in Appendix~\ref{allfigures} 
Figs.~\ref{ColdGuillotSpectra}~a,~b,~\ref{targetsSpec}~a~and~b), 
which is treated differently, depending on the code: 
\emph{petitCODE} uses the linelist from HITRAN 
\citep{Rothman_2013a}, while \emph{ATMO} and \emph{Exo-REM} use 
data from Exomol \citep{Yurchenko_2014}.

There are also some differences between spectra 
(Fig.~\ref{ColdTrans} ) at $\sim$~6~$\mu$m and 10~$\mu$m (and, in 
Appendix~\ref{allfigures}, Figs.~\ref{targetsSpec}~a~and~b) which 
are attributed to the use of different NH$_3$ linelists : 
\emph{petitCODE} uses HITRAN, whereas the other codes use Exomol 
\citep{Yurchenko_2011}. 
The use of only the main CO isotopologue buy \emph{petitCODE} is at 
the origin of the differences observed between 4 and 
5 microns in the high temperature cases (Fig.~\ref{EmissTest} and, 
in Appendix~\ref{allfigures}, 
Figs.~\ref{HotGuillotSpectra}~b~and~c,~\ref{targetsSpec}~c).

The use of different linelists also affects the integral flux of 
the spectrum, i.e. the effective temperature (e.g., in 
Appendix~\ref{allfigures}, the temperature profiles of the two 
self-luminous exoplanets,  Fig.~\ref{targetsPT}, are a bit shifted 
toward lower temperature for \emph{petitCODE}, to compensate the 
lack of absorption). 

These results point out the importance of updating line lists in 
the models in the JWST era.\\

At 2000~K and much more at 2500~K (Fig.~\ref{HotEmis} and, in 
Appendix~\ref{allfigures}, Fig.~\ref{HotGuillotSpectra}~c), \emph{Exo-REM} has more 
flux than the other models. The reason for this is the range of 
validity of this model. Not designed for irradiated objects, 
correlated-k coefficients of \emph{Exo-REM} are not computed for 
high temperature / low pressure. A temperature higher than 1800~K, 
at high altitude (for a pressure $<$ 10$^{-5}$ bar), is out of the 
range of validity of \emph{Exo-REM}.\\ 

Note that, we do not find differences coming from the fact that 
\emph{petitCODE} uses air broadening, or from the correlated-k 
approaches of \emph{ATMO}.

\subsubsection{Differences due to chemistry}

While there is no large difference in the spectra, large 
differences in some molecule and atom abundances can be observed 
(See Fig.~\ref{1000KProf}), especially for PH$_3$ up to 0.1 bar 
for the case at 500~K (Fig.~\ref{1000KProf}~a, and in 
Appendix~\ref{allfigures} 
Figs.~\ref{targetsAbund}~a~and~b), and for the alkali in 
general at 1000~K (Fig.~\ref{1000KProf}~b for 30 solar 
metallicity, and, in Appendix~\ref{allfigures} 
Figs.~\ref{ColdGuillotAbund}~b,~c,~and~d). 
These differences occur each time the condensation curve of a 
compound is crossed, and \emph{Exo-REM} gives systematically lower 
abundances values. This effect is explained by the implementation 
of the cold trap in \emph{Exo-REM}. Indeed, \emph{Exo-REM} makes 
the following assumption: if a molecule condensates at a given 
level in the atmosphere, the material is lost for the upper 
layers. For instance, as alkali condensation occurs deep in the 
atmosphere, \emph{Exo-REM} predicts a lower amount of Na and K 
than \emph{ATMO} and \emph{petitCODE}. These differences do not 
have a significant effect on spectra (Fig.~\ref{ColdTrans}).

\subsection{Convergence process step-by-step \label{stepbystep}}

To arrive at the level of similarity presented in the previous 
section we had to make several changes in our models. In 
this section, we describe how we updated our models and the 
benchmark protocol.\\

Before the beginning of the benchmark, \emph{Exo-REM} was updated 
by adding CO$_2$ to the absorbing molecules list and by updating 
the H$_2$--He CIA sources. \emph{Exo-REM} also implemented a new 
way to combine correlated-k coefficients (described in 
Appendix~\ref{upexo}).\\

When we began this benchmark, we identified a major difference 
between the models concerning how the alkali far wings 
were accounted for (see Sect.~\ref{alkali}). Without any good 
answer to this question (see Sect.~\ref{alkali}), we decided to 
consider the same simple alkali treatment (same Voigt profiles 
with cut-off). 

The other major difference was identified as coming from 
the molecular far wing lineshape. In fact, \emph{petitCODE} did 
not initially include any cut-off in the line profile (unlike 
\emph{ATMO} and \emph{Exo-REM}). Across the full considered 
wavelength range, all lines had a Voigt profile 
\citep[see][Appendix A for the scheme]{Molli_re_2015}.
The first consequence of this comparison is that \emph{petitCODE} 
now also includes a line cut-off as its default line opacity 
treatment, implemented in the same way as in \emph{Exo-REM}. 
Moreover, we study the effect of considering a cut-off, when 
compared to the case without any cut-off application, in 
Section~\ref{cutoff}.

These modifications lead to a significant improvement of the 
convergence. Then, we spotted a strong absorption in 
\emph{petitCODE} results (between 4 and 5~$\mu$m), not visible in 
the results from other models. The corresponding absorber was 
PH$_3$, not added at this stage neither in \emph{ATMO} nor 
\emph{Exo-REM}. This implied an important modification of 
\emph{ATMO} and \emph{Exo-REM} to add the chemistry and absorption 
of phosphine. This molecule has now been included in these models.\\

At this step, spectra obtained with the three models began to show 
good agreements, but we still observed differences that were 
identified as due to methane features and to H$_2$-He CIA. 
These differences came from the use of different CH$_4$ line lists 
and continuum collisional-induced absorption coefficients. 
Indeed at that time \emph{petitCODE} did not use the latest 
version of CH$_4$ linelist but used that of \cite{Rothman_2013}; 
it also was not using the latest CIA values but those of 
\cite{Borysow_2001,Borysow_2002}. \emph{ATMO} used an updated 
version of H$_2$-He CIA \citep{Richard_2012} but not a complete 
one compare to \emph{Exo-REM} (see Appendix~\ref{upexo}). 
Once each model used the CH$_4$ lines list, and the same H$_2$-He 
CIA as \emph{Exo-REM}, the differences disappeared.\\

Three other effects have been identified as being important 
to converge on the results. 

In the extreme case with a T$_\mathrm{eff}$= 2500~K 
\emph{petitCODE} showed a strong effect due to ionization, a 
process not taken into account by the other models; hence we 
decided to deactivate it for the benchmark protocol.

The next two effects have an impact on transmission spectra and 
were observable only at the end of the whole convergence process.
They are taken into account to achieve the remarkably similar 
spectra that we obtained for the self-consistent transiting 
targets WASP~12~b and GJ~436~b (see in Appendix~\ref{allfigures} 
Fig.~\ref{targetsSpec} c).

First, the mean molecular weight (MMW) is known to have a 
strong effect on transmission spectra 
\citep{Miller_Ricci_2009, Croll_2011}. In the beginning of the 
benchmark study, the chemistry code written for and used for the 
first time in \emph{petitCODE} \citep{Molli_re_2017}, erroneously 
included condensate species for calculating the MMW, for which the 
condensate mass fraction and condensate monomer mass 
was used. This mistake was spotted quickly during our benchmark 
process, and corrected before the publication of 
\cite{Molli_re_2017}, and therefore never affected any published 
results and models of \emph{petitCODE}.

The reason for the difference to occur is that the monomers can be 
significantly more massive than the gas species molecules from 
which they are constructed, e.g. for silicates such as 
Mg$_2$SiO$_4$, leading to an overestimation of the MMW, which 
resulted in noticeable differences in the output transmission 
spectra between \emph{ATMO} and \emph{petitCODE}. In reality, 
condensible species only affect the MMW by depleting heavier 
molecules and atoms from the gas phase, hence leading to a 
decrease in MMW, rather than an increase. As the condensate 
particles in the atmosphere, which have masses much larger than 
the monomer mass, will usually not be sufficiently well 
collisionally coupled to the atmospheric gas, they should not be 
taken into account in the calculation of the MMW.

Second, transit spectroscopy is also sensitive to the 
pressure-radius profile used. It appears that we need to have 
exactly the same radius at a pressure of 10~bars to obtain similar 
transmission spectra without an offset.

\section{Focus on key parameters effects \label{keysSect}}

During the benchmark we realized that the uncertainties in two 
inputs of the models, i.e. the alkali line profiles, and the 
molecules far line-wing shape, lead to significant uncertainties 
in the results. In this section we detail the effects of these 
uncertainties.

\subsection{Alkali line profiles \label{alkali}}

At temperatures below the condensation of TiO/VO ($<$2000~K), 
the opacity in the visible part of the spectrum is dominated by 
the wings of the Na-D (5890~\AA) and KI (7700~\AA) resonance 
lines. While an accurate treatment of the line shape of usual 
atomic lines is only needed at $\sim$~25~\AA ~from the line center, 
the alkali resonant lines far wings profile at thousands of 
\AA ngstr\"om from the line center still induce an important 
absorption effect. In the past literature from brown dwarfs 
atmosphere modelling, two groups have tried to address the issue 
of the shape of the alkali-resonant-line far wings:
\begin{itemize}
\item \citet{Burrows_2003} have used standard quantum chemical 
codes and the unified Franck-Condon model in the quasi-static 
limit to calculate the interaction potentials and the wing shapes 
for the alkali resonant lines in H$_2$/He atmospheres.
\item \citet{Allard_2007} have used a unified line shape semi-
classical theory \citep{1999PhRvA..60.1021A} using molecular 
potentials computed using a valence pseudo-potential \citep[see][]
{Allard_2007}.
\end{itemize}

To our knowledge, these two works have never been benchmarked in 
the astrophysics literature. In this section, we illustrate the 
effects of the different alkali treatments by using the 
\emph{ATMO} code on the Guillot--1500K equilibrium PT profile, and 
on the radiative/convective solution for GJ~504~b. We have used 
three treatments:
\begin{itemize}
\item The Voigt profiles as defined in Sect.~\ref{mincore}
\item The "Burrows" profiles as implemented by \citet{Baudino_2015}
\item The "Allard" profiles as used in \citet{Tremblin_2015}
\end{itemize}

\subsubsection{Guillot profile T$_\mathrm{eff}$ = 1500 K}

\begin{figure}[ht]
\begin{center}
\includegraphics[width=1.00\columnwidth]{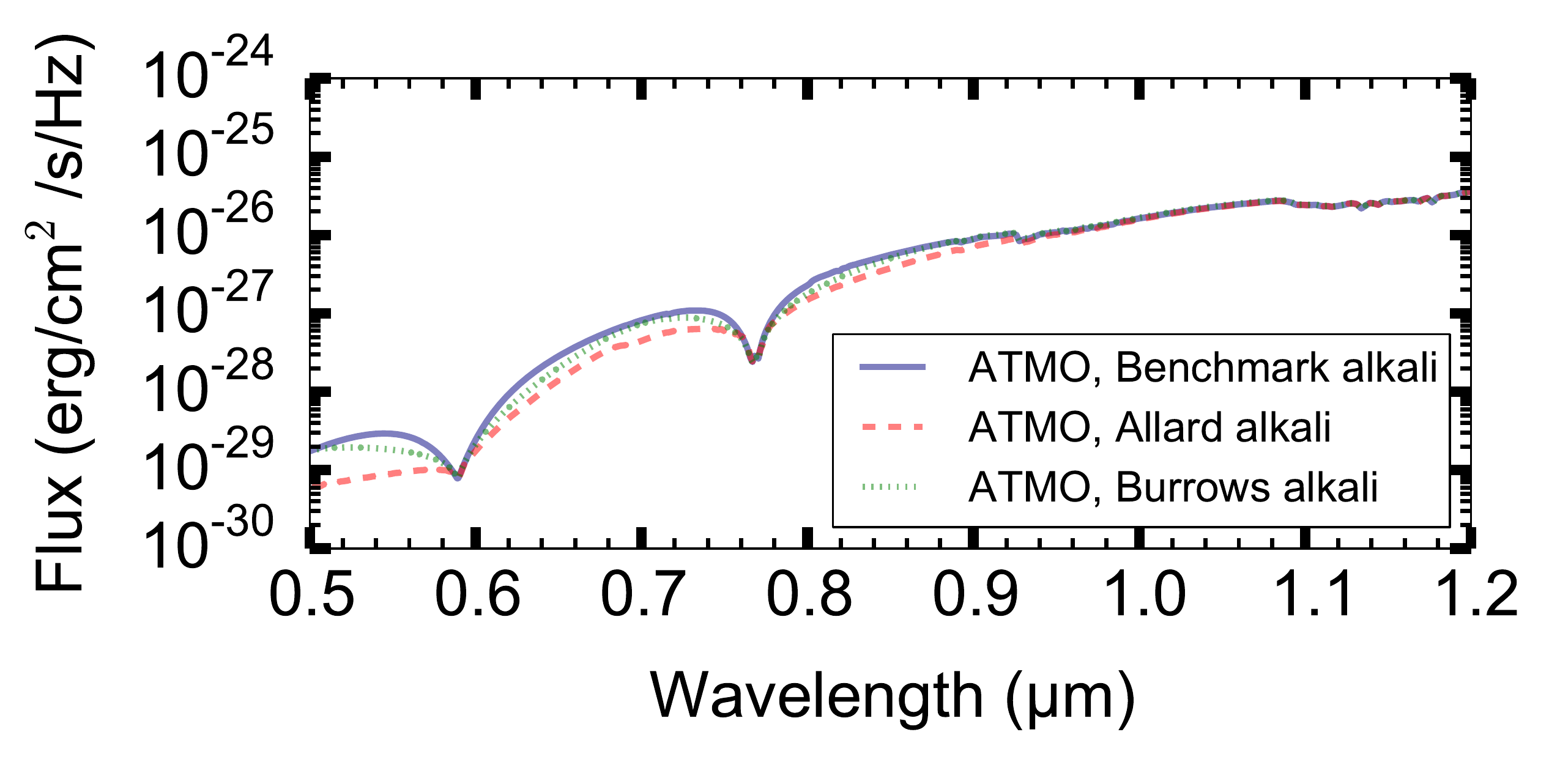}
\caption{{Effects of the different alkali treatments on the emission spectra for the Guillot profile with $T_\mathrm{eff}$=1500K.\label{fig:alkali1}}}
\end{center}
\end{figure}

\begin{figure}[ht]
\begin{center}
\includegraphics[width=1.00\columnwidth]{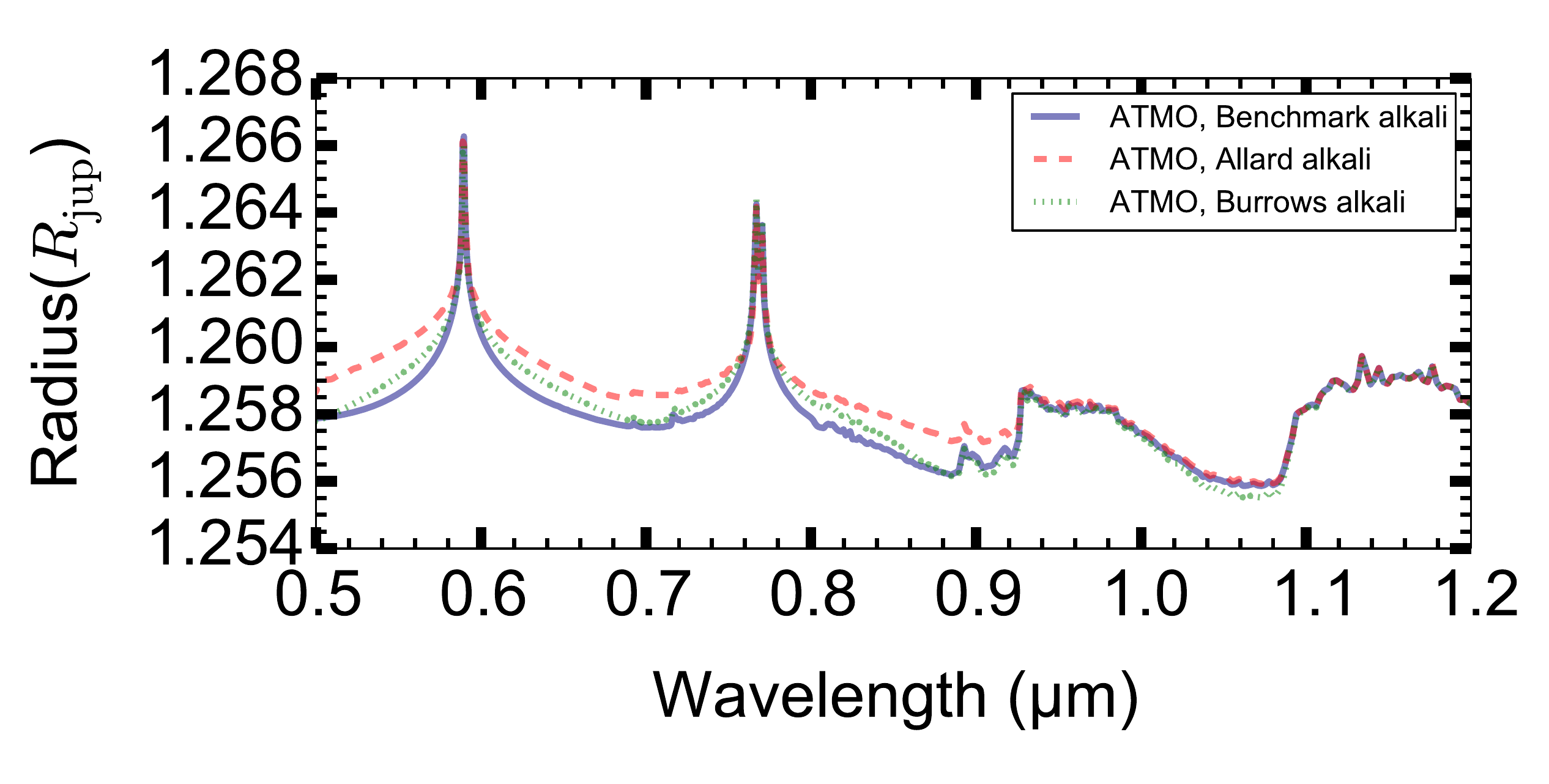}
\caption{{Effects of the different alkali treatments on the transmission spectra for the Guillot profile with $T_\mathrm{eff}$=1500K. \label{fig:alkali2}}}
\end{center}
\end{figure}

We first perform a benchmark study on a fixed PT profile, the 
Guillot T$_\mathrm{eff}$=1500~K profile used in Sect.~2.1. Figures 
\ref{fig:alkali1} and \ref{fig:alkali2} show the emission and 
transmission spectra. Both spectra highlight a low absorption in 
the wings of the Voigt profile; both Burrows and Allard profiles 
increase the absorption in the far wings. Yet, this increase is 
significantly bigger with the Allard profile and the use of one or 
the other is therefore not neutral. For a Jupiter like exoplanet 
transiting a solar type star, the difference between the 
transmission spectrum using Allard profiles or Burrow profiles is 
found to be in the 20~ppm range, which is accessible to 
observations with the JWST. But these differences should be 
attenuated when using models with clouds, so that it might be 
difficult to disentangle which profile fits better the 
observations.

\subsubsection{GJ504b}

\begin{figure}[ht]
\begin{center}
\includegraphics[width=1.00\columnwidth]{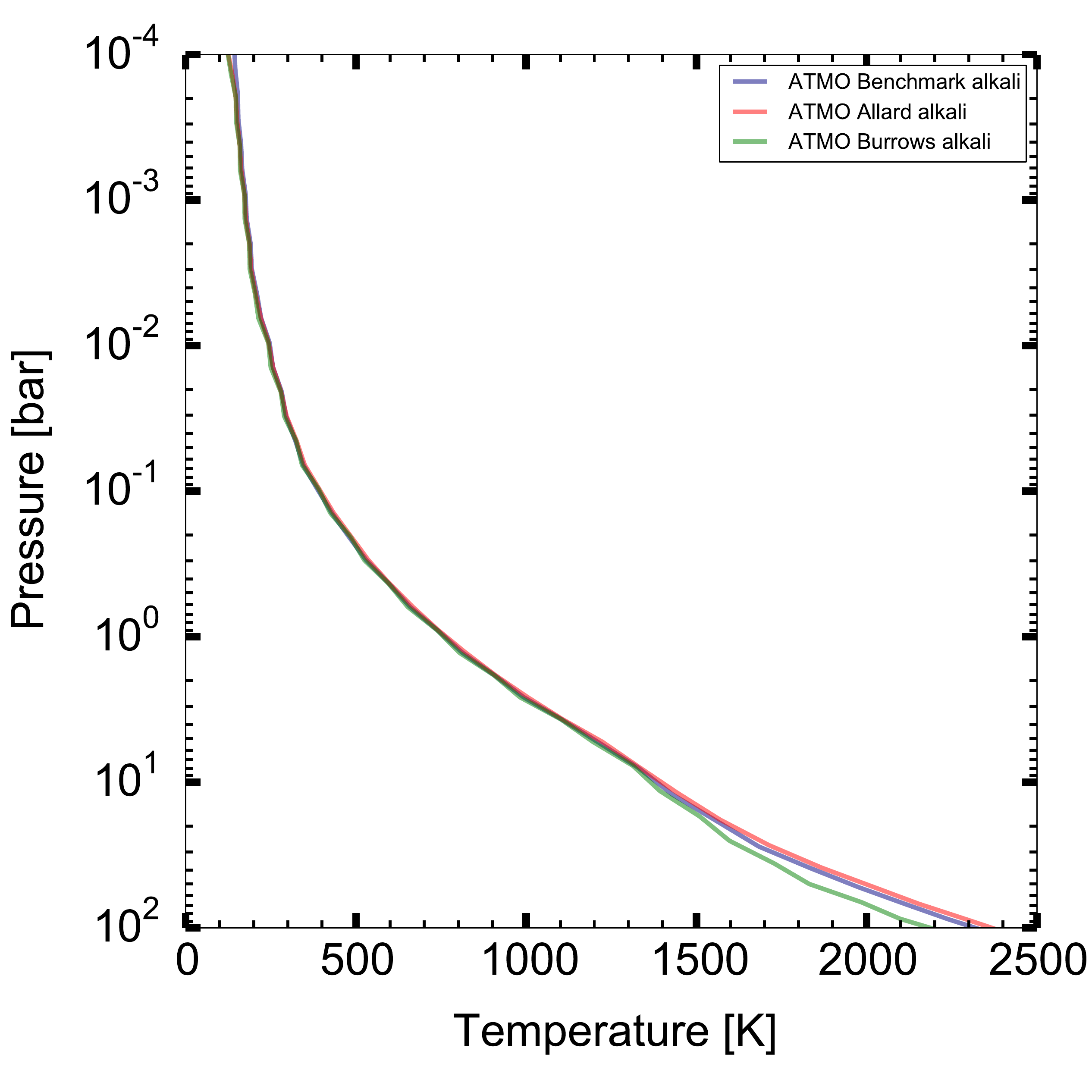}
\caption{{Effects of the different alkali treatments on the PT profiles computed in radiative/convective equilibrium for GJ504b. \label{fig:alkali3}}}
\end{center}
\end{figure}

\begin{figure}[ht]
\begin{center}
\includegraphics[width=1.00\columnwidth]{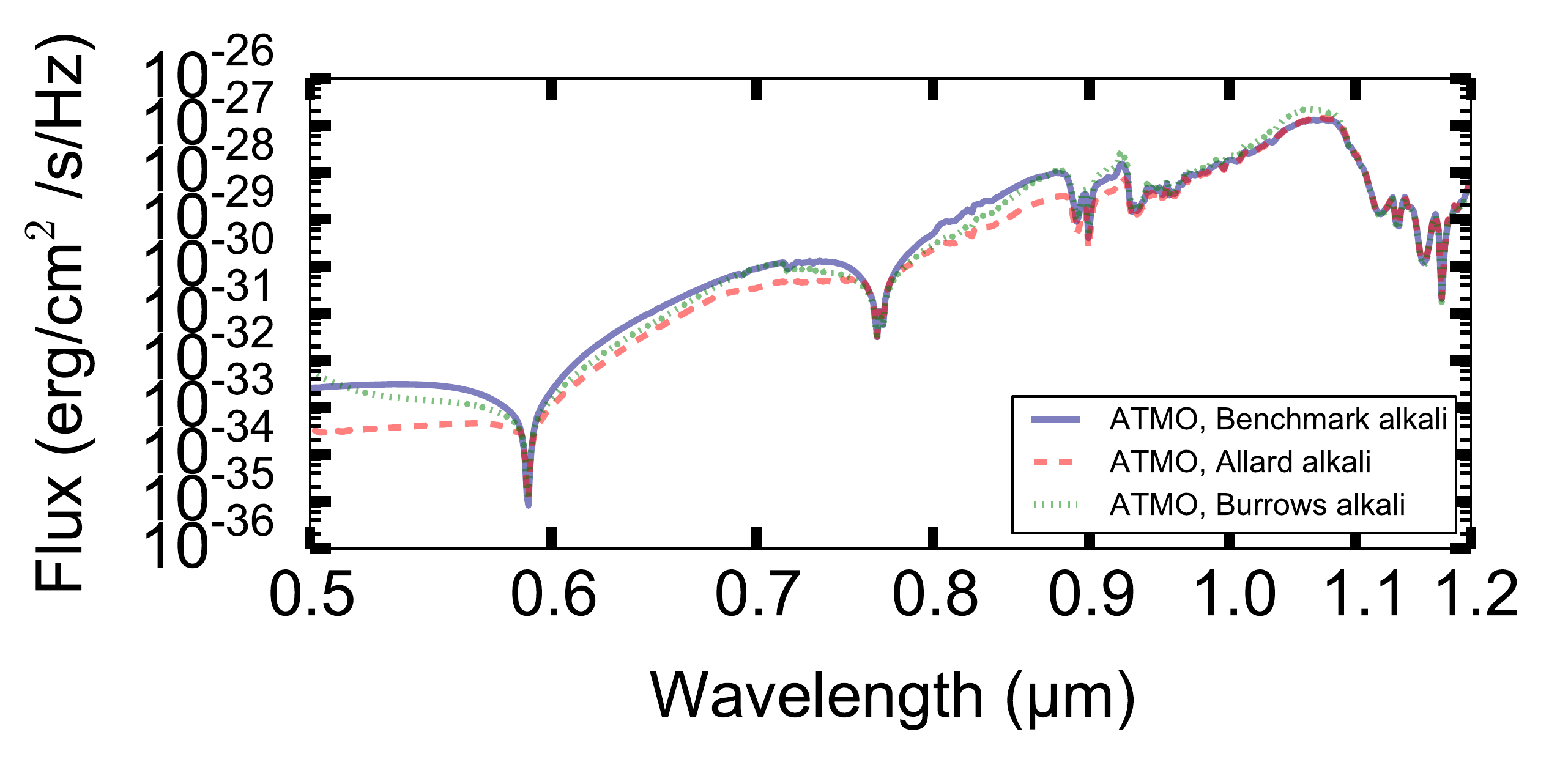}
\caption{{Effects of the different alkali treatments on the emission spectrum of GJ504b. \label{fig:alkali4}}}
\end{center}
\end{figure}

We have used the different alkali treatments in the 
radiative/convective equilibrium model for GJ~504~b. Figure 
\ref{fig:alkali3} shows the effects on the converged PT profile, 
the Allard profile leads to a slightly warmer deep atmosphere than 
the Voigt profile and the Burrows profile to a significantly 
cooler deep atmosphere, with a difference of 100-200~K at 
100~bars. These effects are directly linked to the differences in 
absorption caused by the line profile. Figure \ref{fig:alkali4} 
shows the effects of the different alkali treatments on the 
emission spectrum of GJ~504~b. 
 
The Allard profile leads to a higher absorption both below 
0.9~$\mu$m and above 1~$\mu$m, compared to a simple Voigt profile. 
The Burrows profile leads to a higher absorption, compared to 
Voigt profile, below 0.9~$\mu$m but a lower one above 1~$\mu$m. In 
fact with the Burrows profile, there is no absorption caused by 
the potassium far wing in the Y band while a strong absorption is 
appearing with the Allard profile.

Comparisons with observations as done in \citet{Tremblin_2015} for 
T dwarfs would favour the Burrows profile for this kind of object. 
The authors have used the Allard profile but they had to assume a 
condensation of potassium in KAlSi$_3$O$_8$ to remove all effects 
of the far wing on the Y band. We emphasize again that this 
conclusion depends on the model used to produce the NIR reddening 
(temperature gradient reduction in the case of 
\citealt{Tremblin_2015}). In the other way, a comparison of the 
profile effects on the spectra of L dwarfs seems to favour the 
Allard profile (private communication B. Burmingham). The limits 
we have faced in this alkali-treatment benchmark strongly advocate 
for a community effort to have a better understanding on the 
far-wing behaviour as a function of pressure and temperature. It 
is also of great importance to get better constraints on the 
different cloud models or alternatives.

\subsection{Molecular far wing lineshape \label{cutoff}}

In this section we estimate the effect of applying a line 
"cut-off" (a sub-Lorentzian far wing profile) in the Voigt profile 
of the molecular lines as described in Sect.~\ref{benchmarkcond}. 
For that we compare to a Voigt profile without cut-off hereafter. 
Doing so, we face a difficulty with CH$_4$. Indeed the huge number 
of CH$_4$ lines made the calculation of CH$_4$ opacities without 
line cut-off infeasible (for this paper). To deal with that 
difficulty, we have estimated the effect of no cut-off in two 
ways: 1) first by considering the effect only on a limited 
wavelength range and with a fixed P-T profile 
(Sec.~\ref{fixePTcut-off}), and then by taking into account no cut 
off on all molecules, but CH$_4$, whose opacity is not considered, 
and doing a self-consistent calculation.

\subsubsection{Comparison at fixed temperature structure \label{fixePTcut-off}}

As a first step, we compare the results from a self-consistent
atmospheric calculation for GJ~504~b as obtained by 
\emph{petitCODE} including the sub-Lorentzian lineshape (cut-off), 
to those with full Voigt profiles, using the same temperature 
structure, however. We use the same set of absorbing species as in 
the baseline case. Due to the extensive line list of CH$_4$ 
\citep{Yurchenko_2014} we calculated the opacity for this molecule 
in the full Voigt case only within 1.1 to 1.4~$\mu$m, and 
neglected all lines outside of this spectral region. Therefore we 
will concentrate on this spectral region for our comparison. This 
spectral interval was chosen because both water and methane have 
an opacity minimum ranging from 1.2 to 1.3~$\mu$m, and a change in 
the line wing continuum should be most noticeable in such regions.

\begin{figure}[ht]
\begin{center}
\includegraphics[width=1.00\columnwidth]{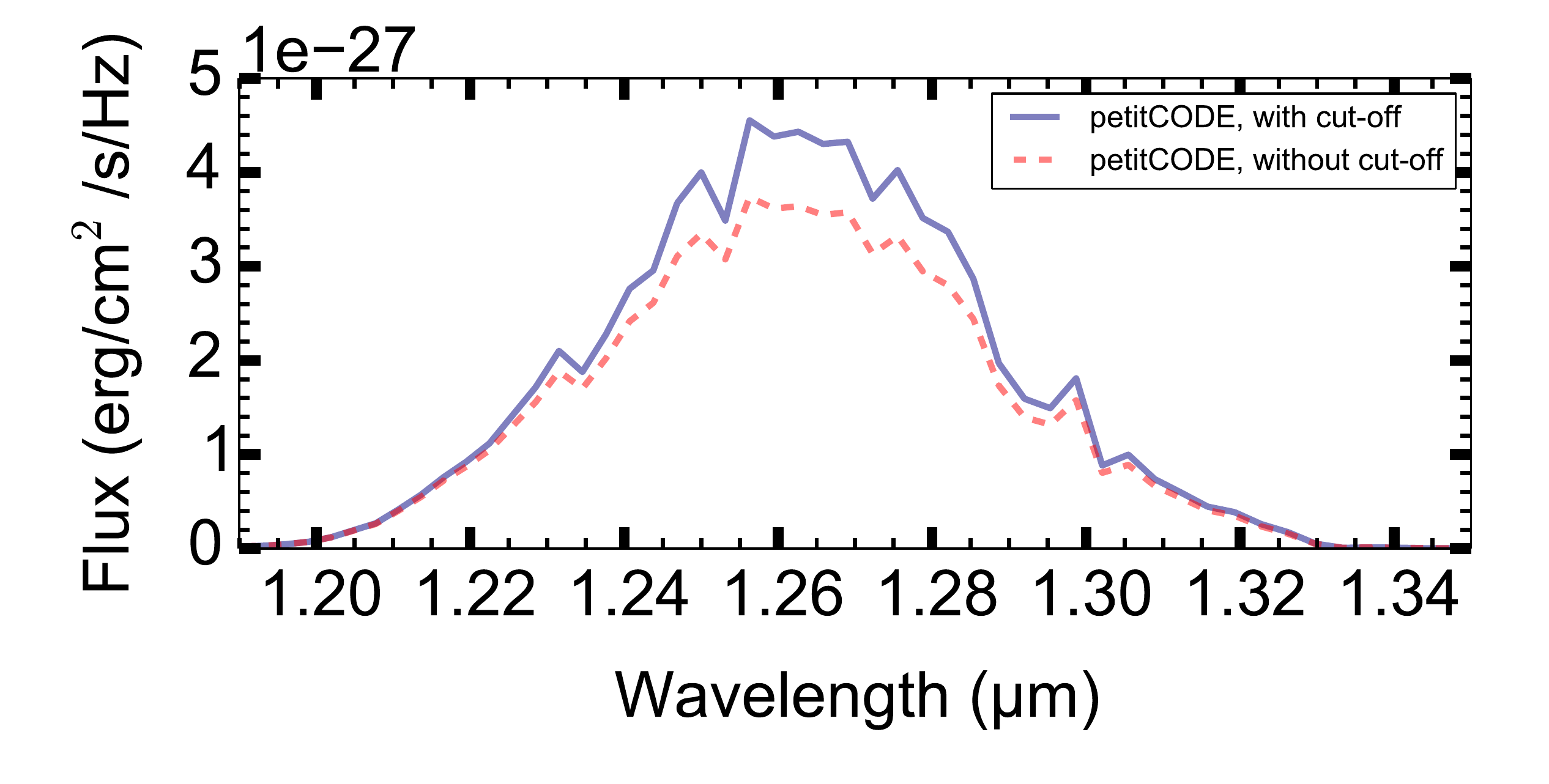}
\caption{{Emission spectra for GJ~504~b calculated using the 
nominal temperature structure of the case including 
sub-Lorentzian far wings. We show the resulting spectrum of the 
nominal case (blue solid line) as well as the case with full Voigt 
profiles (red dashed line). \label{fig:cutoff_emis}}}
\end{center}
\end{figure}

The resulting emission spectra can be seen in 
Figure~\ref{fig:cutoff_emis}. The difference between the two cases 
is strongest in the peak of emission, i.e. in the location where 
the total line opacity was the smallest and the contribution of 
the line wing continuum the largest. We find maximum differences 
in the flux of the order of $\sim 20$~\%. If we had included 
methane lines outside of the 1.1 to 1.4~$\mu$m region the decrease 
in emission may have been even stronger, with lines further away 
potentially still non-negligibly contributing to the line wing 
continuum.

\subsubsection{Comparison for self-consistent structures \label{selfCONcut-off}}

\begin{figure}[ht]
\begin{center}
\includegraphics[width=1.00\columnwidth]{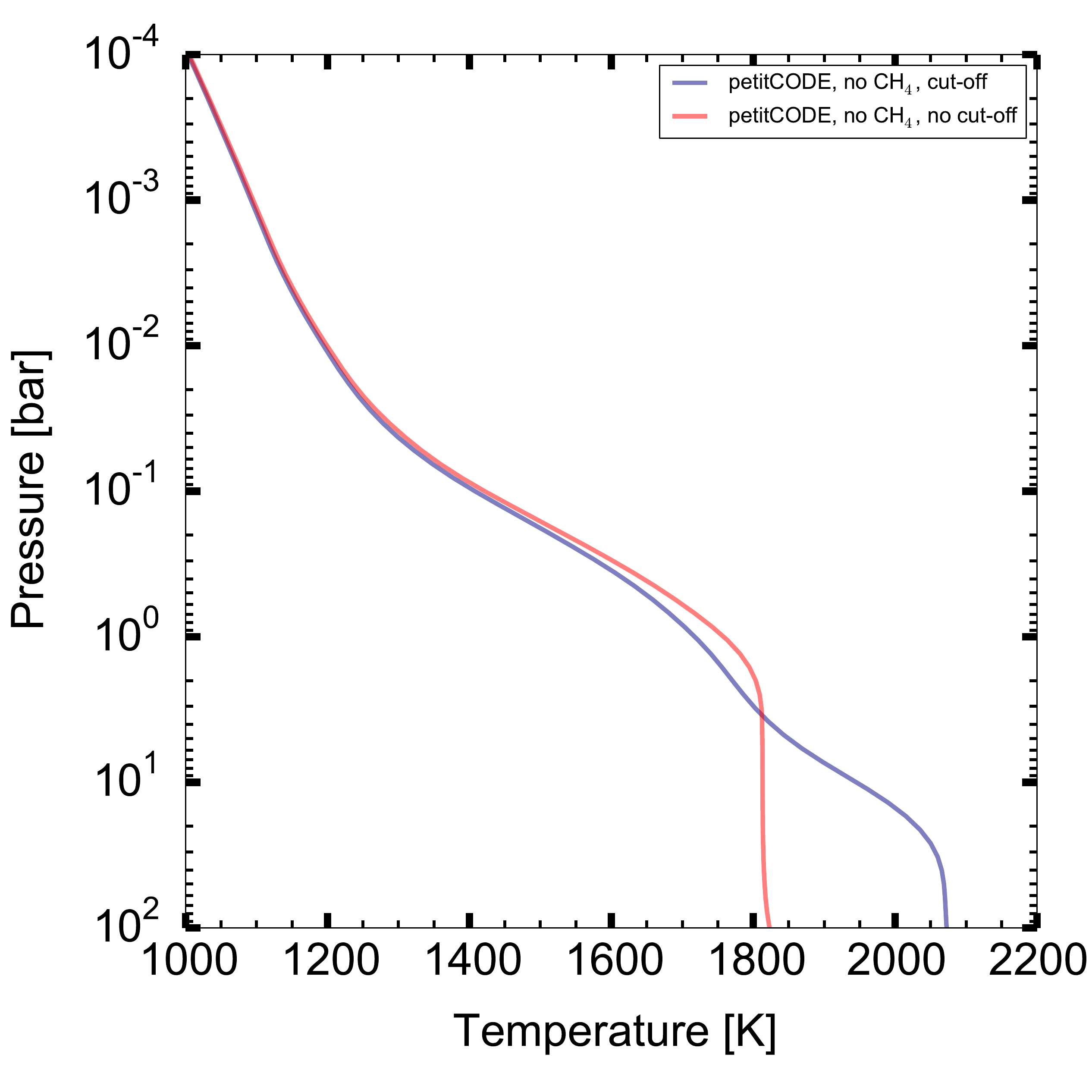}
\caption{{Pressure-temperature structures of the self-consistent 
hot Jupiter cases sub-Lorentzian and Lorentzian far wings, 
neglecting the methane opacity. \label{fig:scPTirrad}}}
\end{center}
\end{figure}

\begin{figure}[ht]
\begin{center}
\includegraphics[width=1.00\columnwidth]{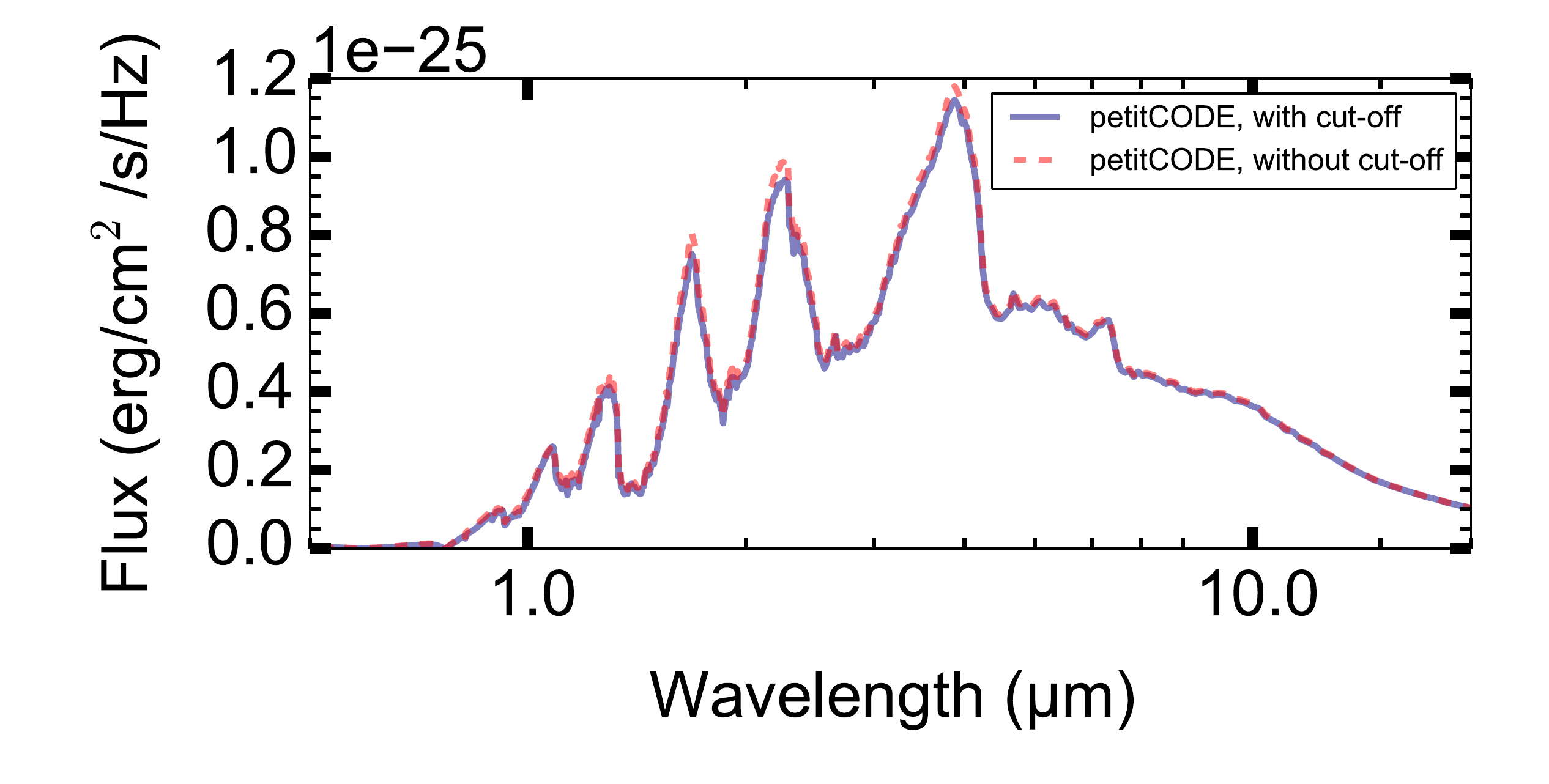}
\caption{{Emission spectra of the self-consistent hot Jupiter 
cases sub-Lorentzian and Lorentzian far wings, neglecting the 
methane opacity. \label{fig:self_cons_emis_irrad}}}
\end{center}
\end{figure}

\begin{figure}[ht]
\begin{center}
\includegraphics[width=1.00\columnwidth]{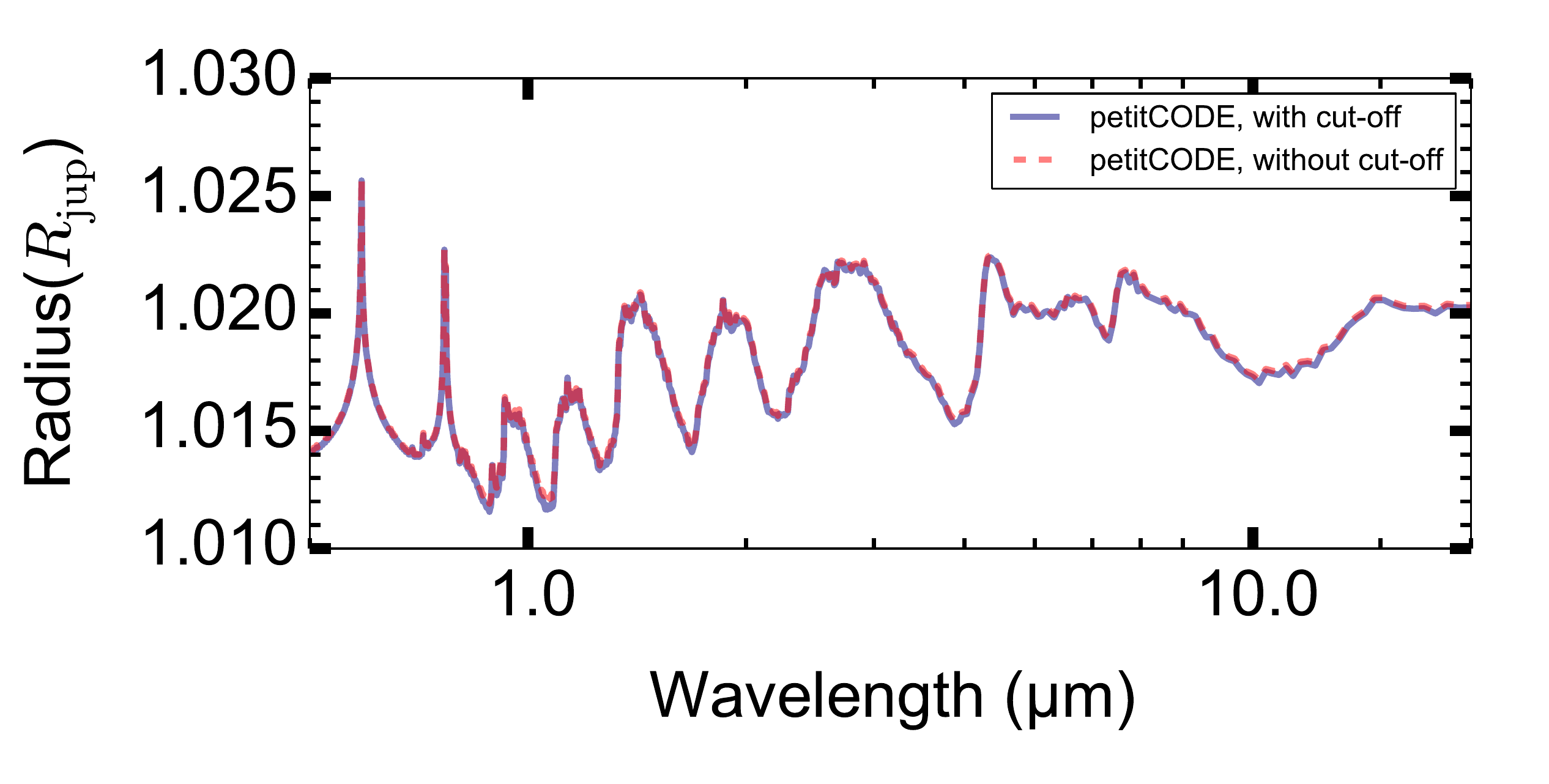}
\caption{{Transmission spectra of the self-consistent hot Jupiter 
cases sub-Lorentzian and Lorentzian far wings, neglecting the 
methane opacity. \label{fig:self_cons_trans_irrad}}}
\end{center}
\end{figure}

The spectral test we carried out in the previous section was done 
only within a small spectral window, due to the extensive CH$_4$ 
line list, which makes the calculation of opacities 
applying a Lorentzian lineshape over the full spectral range 
infeasible and did not allow for comparing self-consistent 
structures and spectra arising from the Lorentzian vs. 
sub-Lorentzian lineshape calculations. In this section we 
therefore vary our baseline model and calculate hot Jupiter 
atmospheric structures without considering CH$_4$ opacities, 
because for all other species an opacity calculation with full 
Lorentzian lineshape is possible. We do not claim that an 
exclusion of CH$_4$ is realistic, although CH$_4$ should not be 
dominant at the elevated temperatures studied here, we merely 
carry out this test to study the effect of a line wing cut-off.
The hot Jupiter is defined as an irradiated planet around a Sun-
like star, with $T_{\rm eff}=1500$~K, $M_{\rm Pl}=1.2 \ {\rm 
M}_{\rm Jup}$, $R_{\rm Pl}=1 \ {\rm R}_{\rm Jup}$ (at 10 bar) and 
$T_{\rm int}=100$~K and an atmospheric enrichment of 3 times 
solar.

The $P$-$T$ structures, emission and transmission spectra of the 
self--consistent hot Jupiter cases with sub-Lorentzian and 
Lorentzian far wing, and neglecting the methane opacity, are shown 
in Figs.~\ref{fig:scPTirrad}, \ref{fig:self_cons_emis_irrad}, 
\ref{fig:self_cons_trans_irrad}.

It can immediately be seen that the deep isothermal region of the 
full Voigt atmosphere is cooler, although the total opacity there 
is higher: The opacity increase arising from pure Voigt profiles 
causes the stellar light to be absorbed already at higher layers, 
leading to a small temperature increase between 0.1 and 3~bar in 
this atmosphere. Consequently, less flux has to be emitted from 
the optically thick, deeper regions of the planet, leading to a 
cooler temperature structure for pressures larger than 3~bar. 
Thus, in the presence of irradiation, the calculated temperature 
structure of an atmosphere can change significantly, depending on 
whether or not a Lorentzian lineshape is applied.

Because the higher regions of the atmosphere are not so strongly 
affected by the line far wings the flux changes in the absorption 
features of the emission spectra are small. The emission peaks, 
with radiation originating in the deeper and hotter regions of the 
atmosphere are affected more strongly, with maximum differences of 
7~\%, e.g. at 2.25~$\mu$m.

Over a wide wavelength range the differences between the two cases 
in the transmission spectrum are small, due to the fact that the 
transmission probes high layers of the atmosphere where (i) the 
temperature of the two cases is very similar and (ii) pressure 
broadening is less important. However, the largest differences in 
the optical and NIR, again at locations of minimum opacity, can be 
as large as $\sim 10$~\% of the transmission signal amplitude 
(e.g. Y-band in Fig.~\ref{fig:self_cons_trans_irrad}). Such 
differences are greater than the expected precision of JWST 
observations of hot Jupiter around bright stars.

\section{Non-equilibrium chemistry \label{nonequilibrium}}
\subsection{Impact of non-equilibrium chemistry}
Known since the 70's in the atmospheres of the giant planets of 
our Solar System \citep{Prinn_1977}, non--equilibrium chemistry 
was discovered in a cool brown dwarf, Gl~229~b, in 1997 
by~\citet{Noll_1997}. More recently, it started to be considered 
for transiting exoplanets \citep[e.g. ][]{Moses2011, Line2013} and 
evidence for non--equilibrium chemistry was also found in 
extrasolar directly imaged planets of the HR~8799 system 
\citep[e. g.][]{Barman_2011,Barman_2015,Konopacky_2013} through 
the concomitant detection of CO absorption features and the 
unexpectedly weak absorption by CH$_4$.\\
   
This important phenomenon, which can strongly affect the 
atmospheric chemical composition, is due to a competition between 
vertical mixing and chemical kinetics.  When the dynamical 
timescale is shorter than the chemical timescale we observe a 
freezing of the abundance of some species. The "quenched" species 
keep the same abundances than that of the deep hot layers, where 
chemical equilibrium is achieved. This phenomenon, called 
transport--induced quenching, affects particularly the CO/CH$_4$ 
and N$_2$/NH$_3$ ratios that can be driven away from what is 
expected from equilibrium chemistry. For the atmospheres highly 
affected by quenching, the interpretation of observational spectra 
can thus be quite difficult.

\subsection{Description of the chemicals models}
To model the non-equilibrium chemistry that takes place in warm 
exoplanet atmospheres, two different approaches can be used: a 
complete modelling of kinetics with a model accounting for 
atmospheric mixing \citep[like in \emph{Venot chemical model}, ][
hereafter \emph{V12}]{Venot_2012}, or a chemical equilibrium model 
coupled to a parametrization of the quenching level for the main 
species (like in \emph{Exo-REM}).\\
            
\emph{ Venot chemical model} \\
The thermo--photochemical model developed by \emph{V12} is a full 
1D time-dependent model. It includes kinetics, photodissociation 
and vertical mixing (eddy diffusion and molecular diffusion). To 
determine the atmospheric composition of a planet, the thermal 
profile is divided in discrete layers ($\sim$100) of thickness 
equal to a fixed fraction of the pressure scale height. The code 
solves the continuity equation (Eq.~\ref{eq:continuite}) for each 
species and for each atmospheric layer, until a steady-state is 
reached.

\begin{equation}\label{eq:continuite}
\frac{\partial n_i}{\partial t} = P_i - n_iL_i - div({\Phi_i}\overrightarrow{e_z})
\end{equation}
where $n_i$ the number density of the species $i$ 
($\mathrm{cm^{-3}}$), $P_i$ its production rate 
($\mathrm{cm^{-3}~s^{-1}}$), $L_i$ its loss rate 
($\mathrm{s^{-1}}$),  and $\Phi_i$ its vertical flux 
($\mathrm{cm^{-2}~s^{-1}}$) that follows the diffusion equation,

\begin{equation}
\Phi_i = -n_iD_i \left[ \frac{1}{n_i}\frac{\partial n_i}{\partial z}+\frac{1}{H_i}+\frac{1}{T}\frac{dT}{dz}\right]-n_iK_\mathrm{zz}\left[\frac{1}{y_i}\frac{\partial y_i}{\partial z}\right],
\end{equation}
where $y_i$ is the mixing ratio, $K_\mathrm{zz}$ is the eddy 
diffusion coefficient ($\mathrm{cm^2~s^{-1}}$), $D_i$ is the 
molecular diffusion coefficient ($\mathrm{cm^2~s^{-1}}$), and 
$H_i$ the scale height of the species $i$.\\

At both upper and lower boundaries, we impose a zero flux for each 
species.

The strength of this model relies on the chemical scheme it uses: 
the C$_0$--C$_2$ scheme. It describes the kinetics of 105 species 
made of H, C, O, and N, which are linked by $\sim$2000 reactions. 
This scheme has been implemented in close collaboration with 
specialists of combustion and has been validated experimentally as 
a whole (not only each reaction individually). The completeness of 
this scheme (both the forward and the reverse directions of each 
reaction are considered) implies that thermochemical equilibrium 
is achieved kinetically, or, conversely, that out of equilibrium 
processes are considered. The temperatures found in warm exoplanet 
atmospheres (in particular where quenching happens) are within the 
very large range of validation of the scheme ([300--2500] K~and 
[0.01--100]~bar), leading to a high level of confidence in the 
results obtained. Because the model is time-dependent, it 
naturally covers the phenomenon of quenching.

It is important to note that, contrary to \emph{Exo-REM}, the 
chemical network of \emph{V12} does not account for P--bearing 
compounds, alkalies and silicates.\\

\emph{ Exo-REM model}\\ 
The way \emph{Exo-REM} accounts for non-equilibrium chemistry of 
the CO--CO$_2$--CH$_4$ and N$_2$--NH$_3$ networks is based on the 
analysis performed by \citet{Zahnle_2014} (hereafter \emph{ZM14}). 
These authors applied a 1-D full kinetics model to a set of 
atmospheric models for self-luminous objects with $T_\mathrm{eff}$ 
between 500 and 1100~K, along with a range of $g$, metallicity and 
vertical diffusivity. The objective was to determine the quench 
conditions that yield the asymptotic abundances of various 
non-equilibrium species in the upper atmosphere and describe those 
with simple equations. These equations provide a so-called 
chemical time ($t_\mathrm{chem}$) expressed as a function of 
temperature, pressure and, in some cases, metallicity (Eqs.~12-14 
for the CO--CH$_4$ system, Eq.~44 for CO--CO$_2$, Eq.~32 for the 
N$_2$--NH$_3$ system in \emph{ZM14}). Equaling $t_\mathrm{chem}$ 
to a dynamical time constant $t_\mathrm{mix}$ defined as 
$H^2$/$K_\mathrm{zz}$, where $H$ is the atmospheric scale height 
and $K_\mathrm{zz}$ the eddy mixing coefficient, determines the 
quench level for the considered chemical system. The asymptotic 
abundances at high atmospheric levels are then given by those at 
this quench level. In \emph{Exo-REM}, the mole fractions are 
simply set to those at thermochemical equilibrium up to the quench 
level, and kept constant above this level. Doing so, we do not 
accurately represent the transition region located around the 
quench level, in which the composition gradually freezes, but 
reproduce the full kinetics model calculations of \emph{ZM14} 
below and above the quench level. Note that, as the CO--CO$_2$ 
system is quenched at lower temperatures and thus higher levels 
than CO--CH$_4$, the sum of the CO and CO$_2$ mole fractions is 
kept constant between these two quench levels while the CO/CO$_2$ 
ratio is still determined by thermochemical equilibrium.
      
In addition to the chemical networks investigated by \emph{ZM14}, 
\emph{Exo-REM} now includes that of P-bearing compounds. In 
Jupiter and Saturn, phosphine (PH$_3$) is observed with mole 
fractions of a few ppmv \citep{Ridgway_1976,Larson_1980}, orders 
of magnitude larger than the chemical equilibrium values 
\citep{Fegley_1994}. Investigating a reaction network involving 
H/P/O compounds, \citet{Wang_2016} concluded that at equilibrium, 
H$_3$PO$_4$ is the major P-bearing species at temperatures below 
700 K in Jupiter and Saturn and identified the limiting step for 
PH$_3$ to H$_3$PO$_4$ conversion. From their kinetic model, they 
concluded that for any reasonable value of $K_\mathrm{zz}$ the 
quench level for this reaction is at $T \approx$900~K, well below 
that where condensation of H$_3$PO$_4$ is expected 
($\approx$700~K), so that PH$_3$ is still the dominant P--bearing 
species at observable levels. We checked that this conclusion also 
applies to young giant exoplanets, which have $T_\mathrm{eff}$ 
much higher than Jupiter and Saturn. For example, in GJ~504~b, the 
quench level is around 1000~K while the H$_3$PO$_4$ condensation 
level is $\approx$520~K. Thus the non-equilibrium scheme of 
\emph{Exo-REM} simply discards formation of liquid H$_3$PO$_4$ and 
removes this compound from the list of P-bearing species used to 
calculate the phosphine vertical profile.

\subsection{Comparison of results from different models \label{comparneq}}

We first compare the two approaches previously introduced and then 
(Sect.~\ref{effectneq}) we show the general effects (independent 
of the model) of non-equilibrium chemistry.\\

In this part we use the same elemental abundances defined 
in the benchmark protocol. To compare the results obtained with 
the two chemical models, we apply the following process. 
First, we generate a temperature profile 
with the radiative-convective equilibrium model \emph{Exo-REM}. 
Then this input profile is used both by \emph{Exo-REM} and 
\emph{V12} to compute the abundance profiles. Finally 
\emph{Exo-REM} performs the radiative transfer and generate 
the corresponding synthetic spectra. Because \emph{V12} does not
consider Na, K and PH$_3$ which are nevertheless important in 
radiative transfer, we used for these three species the abundances 
computed by \emph{Exo-REM} in all spectra. We do this work for 
the two directly imaged planets and for the irradiated planet 
with $T_\mathrm{eff}$=1000~K.

\begin{figure}[ht]
\begin{center}
\includegraphics[width=1.00\columnwidth]{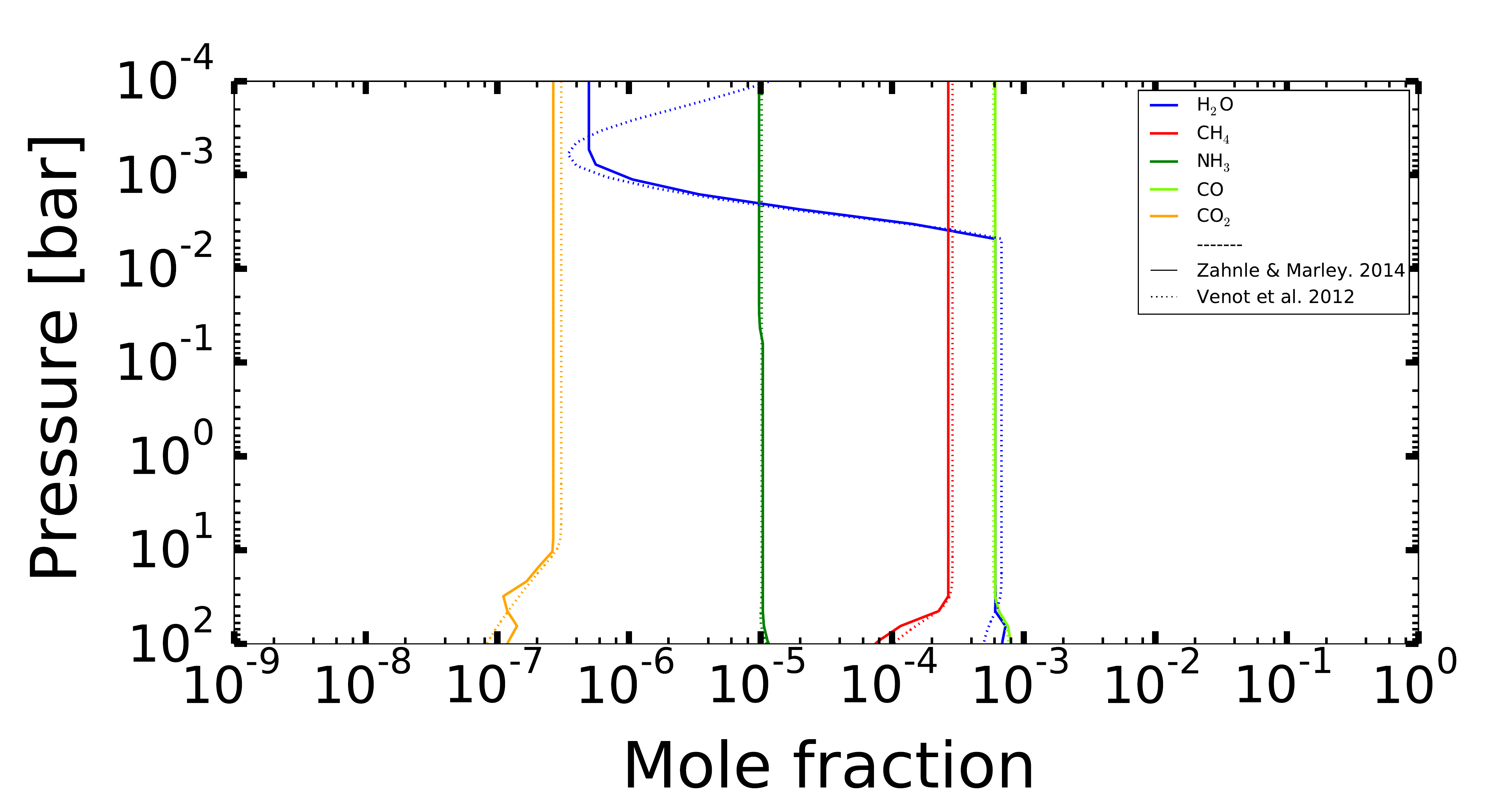}
\caption{{Abundance profiles of the defined molecules for the case 
GJ~504~b and K$_{zz} = 10^{11} $cm$^2 $s$^{-1}$ \label{GJ504b1e11}. 
Solid lines correspond to the simplified approach of 
\emph{Exo-REM} while dotted lines correspond to the chemical 
kinetic model of \emph{V12}.}}
\end{center}
\end{figure}

\begin{figure*}[ht]
\begin{center}
\includegraphics[width=1.\textwidth]{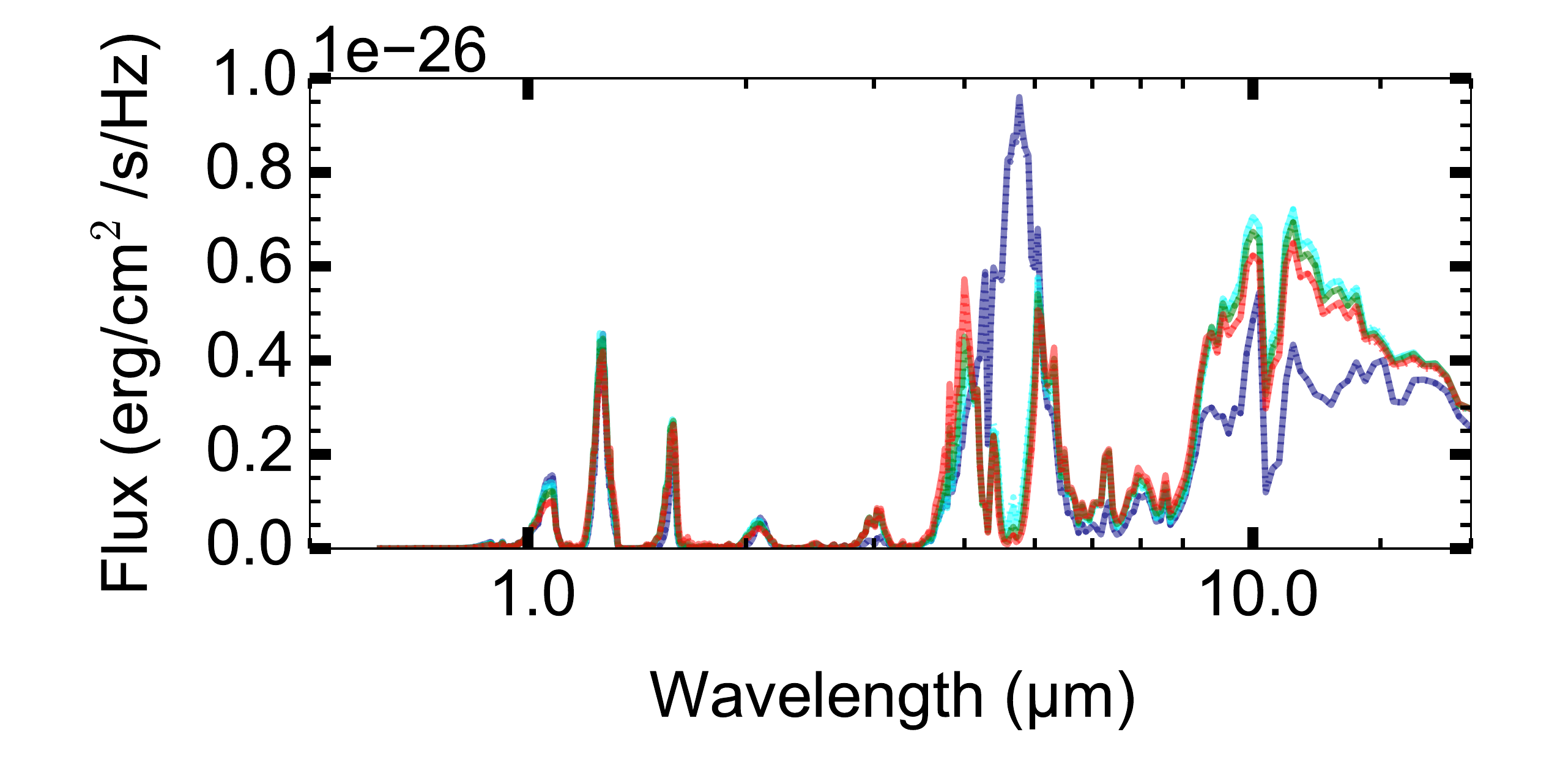}
\caption{{Emission spectra of GJ~504~b considering an atmosphere 
at chemical equilibrium (dark blue) and at disequilibrium (
K$_\mathrm{zz}$ = 10$^{7}$, 10$^{9}$ and 
10$^{11} $cm$^2 $s$^{-1}$, respectively cyan, green and red). 
Spectra have been calculated using the chemical composition 
determined by the kinetic model of \emph{V12} (dotted lines)
and \emph{Exo-REM} (solid lines).
\label{GJ504bEmissNeq}}}
\end{center}
\end{figure*}

\begin{figure}[ht]
\begin{center}
\includegraphics[width=1.00\columnwidth]{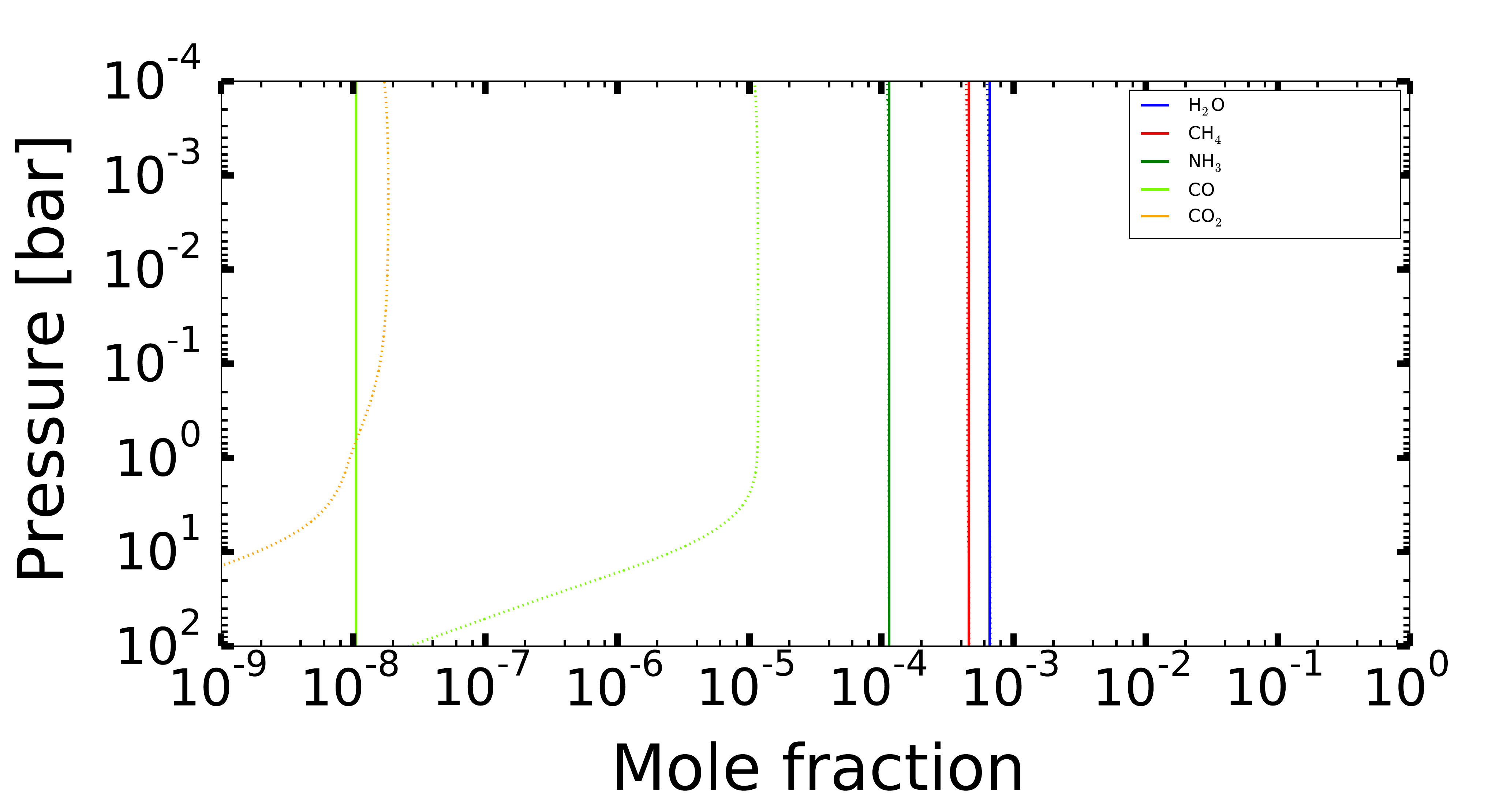}
\caption{{Abundance profiles for the case at 1000 K and 
K$_{zz} = 10^{7} cm^2 s^{-1}$ \label{adund1000K1e7}. Solid lines correspond to the 
simplified approach of \emph{Exo-REM} while dotted lines correspond to the chemical 
kinetic model of \emph{V12}.}}
\end{center}
\end{figure}

\begin{figure}[ht]
\begin{center}
\includegraphics[width=1.00\columnwidth]{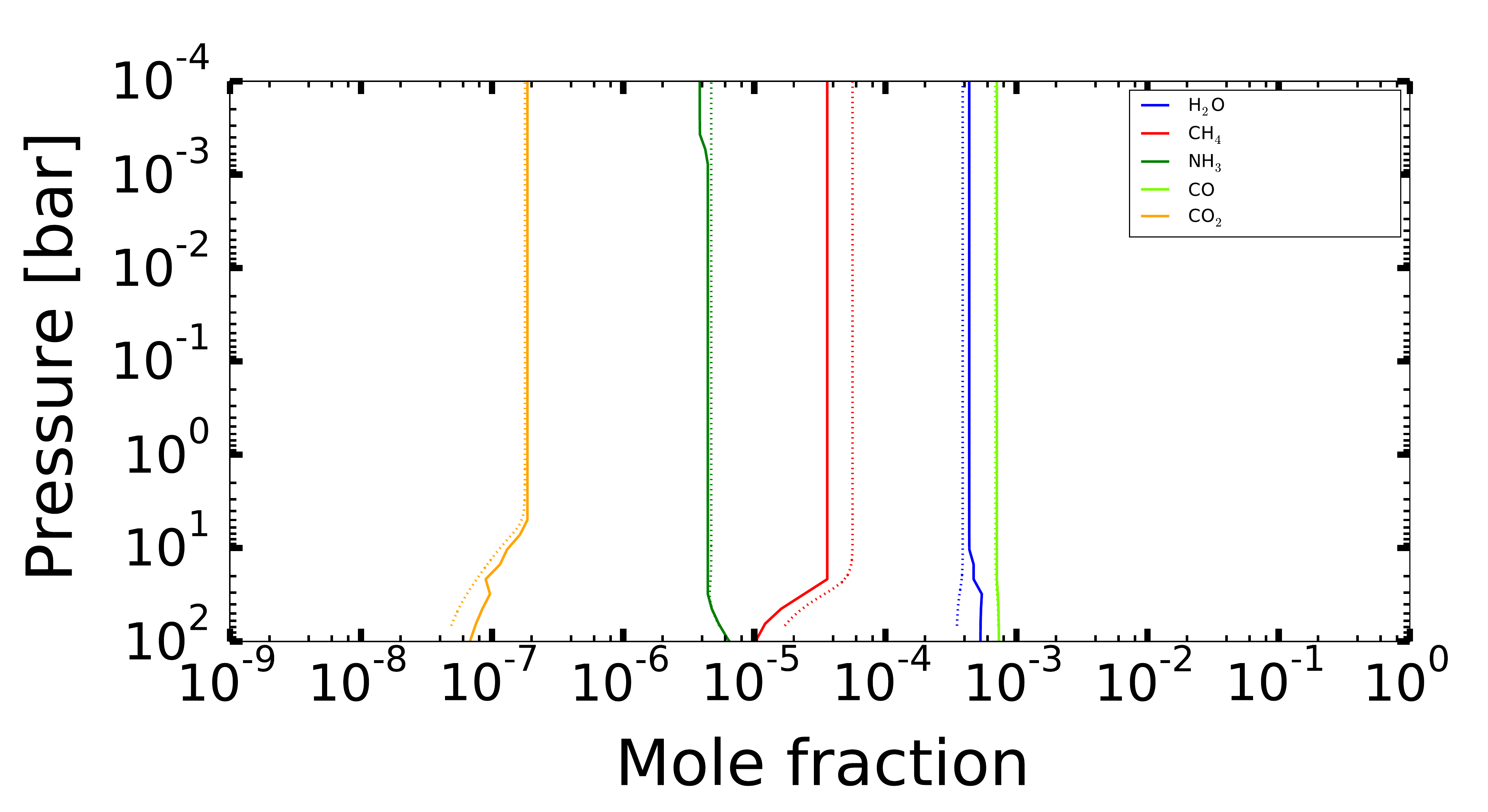}
\caption{{Abundance profiles of the defined molecules for the case 
VHS~1256--1257~b and K$_{zz} = 10^{11} $cm$^2 $s$^{-1}$  
\label{abundVHSbkzz1e11}. Solid lines correspond to the simplified 
approach of \emph{Exo-REM} while dotted lines correspond to the 
chemical kinetic model of \emph{V12}.}}
\end{center}
\end{figure}

We begin this comparison with the thermochemical equilibrium 
composition. In all simulated cases at equilibrium, the abundances 
profiles obtained with the two codes are very similar. However, we 
observe some differences for H$_2$O, CO, CH$_4$ (about a factor of 
2 for CO and CH$_4$) in the deep atmosphere due to the silicate 
condensation taken into account in \emph{Exo-REM} but not in the 
\emph{V12} model. To overcome this difference, we decided to apply 
a modification of the initial elemental abundances used in 
\emph{V12}, in all the cases of this section, corresponding to a 
sequestration of $\sim$20\% of the amount of oxygen in silicates 
(see in Appendix~\ref{allfigures}, 
Figs.~\ref{1000KProf}~a,~\ref{GJ504neqAbund}~a,~\ref{abundVHSneq} 
are after the correction).\\

For the non-equilibrium cases, the abundance profiles show us that 
for the two self-luminous targets we have a good agreement. But 
this is not the case for the atmospheric model corresponding to 
T$_\mathrm{eff}$=1000~K (Fig.~\ref{adund1000K1e7}, and, in 
Appendix~\ref{allfigures}, Fig.~\ref{1000neqAbund}~b-d). In fact 
we observe a strong difference in CO and CO$_2$ treatment (we find 
1000$\times$ more CO with \emph{V12}), coming from the fact that 
\emph{ZM14}, designed for self-luminous planets, do not address 
the kinetic inhibition against oxidizing CH$_4$ (we find 18$\%$ 
less CH$_4$ with \emph{V12}) to CO contrary to \emph{V12}.\\

\begin{figure*}[ht]
\begin{center}
\includegraphics[width=1.\textwidth]{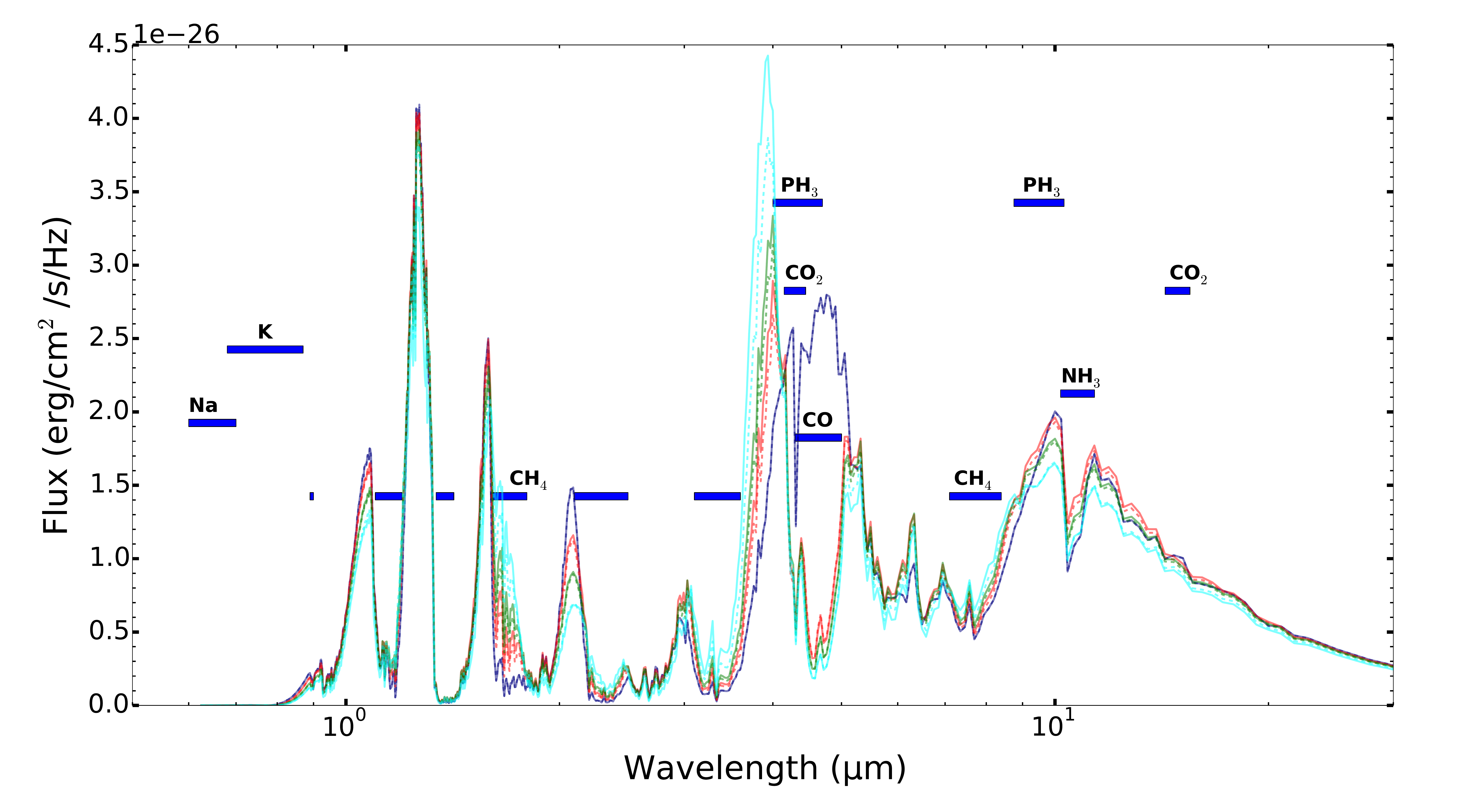}
\caption{{Emission spectra of VHS~1256--1257~b at chemical 
equilibrium (dark blue), or in non-equilibrium chemistry 
(K$_\mathrm{zz}$ = 10$^{7}$, 10$^{9}$ and 10$^{11}$ cm$^2$ 
s$^{-1}$, respectively cyan, green and red) with our two chemical 
models. Location of molecular features, other than H$_2$O, are 
indicated\label{VHSbEmissNeq}. Solid lines correspond to the 
simplified approach of \emph{Exo-REM} while dotted lines 
correspond to the chemical kinetic model of \emph{V12}.}}
\end{center}
\end{figure*}

Other specific and limited differences arise from different 
physical assumptions in the two models. For GJ~504~b,  up to an 
altitude corresponding to 8~mbar there is a clear effect of water 
cold trap (see description of \emph{Exo-REM} in 
Sect.~\ref{modelsDesc}) with \emph{Exo-REM}, not seen in 
\emph{V12} where it is not used (Fig.~\ref{GJ504b1e11} and, in 
Appendix~\ref{allfigures}, Fig.~\ref{GJ504neqAbund}). In the other 
side at the top all atmosphere at $K_\mathrm{zz}$=$10^7$ cm$^2$ 
s$^{-1}$ the difference comes from molecular diffusion, not 
implemented in \emph{Exo-REM} (this difference appears at 
pressures lower than shown in the figures). Condensation of NH$_3$ 
is also a difference, \emph{Exo-REM} takes into account the 
possibility of NH$_4$Cl and NH$_4$SH formation, explaining the 
difference at 8~mbar for VHS~1256--1257~b and 40~mbar for GJ~504~b 
(\emph{V12} does not consider these species, 
Figs.~\ref{GJ504b1e11},~\ref{abundVHSbkzz1e11} and, in 
Appendix~\ref{allfigures}, 
Figs.~\ref{GJ504neqAbund}~b-d,~\ref{abundVHSneq}~b-d).\\

While not obvious for GJ~504~b (Fig.~\ref{GJ504b1e11}), for 
VHS~1256--1257~b (Fig.~\ref{abundVHSbkzz1e11}) we see a difference 
coming from the fact that the \emph{V12} approach exactly computes 
the quenching process (giving a realistic transition across the 
region of chemical quenching) when compared to \emph{Exo-REM} 
where the quenching level is imposed at one precise location. We 
observe more water (10~$\%$) and less methane (50~$\%$) up to 1 bar 
in \emph{Exo-REM}.\\

To conclude, for directly imaged planets the two chemical 
approaches give similar results, but this comparison confirms that 
\emph{Exo-REM}, which uses the approach of \emph{ZM14}, is not 
adapted to study the chemical composition of irradiated planets.\\

\emph{Effect on emission spectra}\\

In this part, we detail the effect of non-equilibrium chemistry, 
on emission and transmission spectra.

Figures \ref{GJ504bEmissNeq} and \ref{VHSbEmissNeq} present 
spectra obtained with the abundance profiles showed in 
Figs.~\ref{GJ504b1e11}, and \ref{abundVHSbkzz1e11}. We can observe 
that the two chemical approaches (\emph{V12} and \emph{Exo-REM}) 
yield very similar spectra. It is difficult to differentiate them  
and the differences are within the error bars expected for 
VHS~1256--1257~b introduced in Fig.~\ref{EmissTest}. High altitude 
departures observed in the abundance profiles (see previous 
subsection) have no visible effect on the spectra because there is 
not enough material to absorb efficiently at pressures lower than 
1~mbar.

\subsection{Intrinsic effect of non-equilibrium chemistry \label{effectneq}}
We use \emph{Exo-REM} in the full benchmark condition except for 
the non-equilibrium chemistry to compute GJ~504~b emission 
spectra, and the full \emph{petitCODE} (no constrained to 
benchmark conditions), combined with \emph{V12} non-equilibrium 
chemistry (and also at equilibrium) for the "Guillot" 1000~K case 
in emission and transmission.\\

Important differences are observed between equilibrium and 
non-equilibrium chemistry at two locations of the spectrum: 
between 4 and 5~$\mu$m, and between 8 and 20~$\mu$m. The first 
location is dominated by absorption of PH$_3$, CO and CO$_2$, in 
out-of-equilibrium chemistry conditions. In the mid-infrared it 
seems easier to discriminate between out-of-equilibrium cases 
regarding $K_\mathrm{zz}$. CH$_4$ and NH$_3$ have the biggest 
impact at this location and the depth of the line of the latter 
near 10~$\mu$m offer a possible way of constraining 
$K_\mathrm{zz}$ (Figs.~\ref{GJ504bEmissNeq}, \ref{VHSbEmissNeq}). 
PH$_3$ impacts also the wavelength range just beyond 10~$\mu$m, 
and at 15~$\mu$m the signature of CO$_2$ appears with 
non-equilibrium chemistry (Fig.~\ref{GJ504bEmissNeq}).\\

The coronagraphic mode of JWST's MIRI instrument has been designed 
to be sensitive to the depth of the NH$_3$ feature, especially 
between the bottom (10.65~$\mu$m) and the edge of the band at 
(11.4~$\mu$m)(Figs.~\ref{GJ504bEmissNeq}, \ref{VHSbEmissNeq}).The 
depth of this feature seems to be a good way to investigate non-
equilibrium chemistry. 

Keeping in mind the fact that temperature profiles of GJ~504~b 
(Fig.~\ref{GJ504bEmissNeq}) and VHS~1256-1257~b 
(Fig.~\ref{VHSbEmissNeq}), for each $K_\mathrm{zz}$, are computed 
self-consistently, we see some differences in the spectral energy 
distribution. This occurs because the model computes the spectrum 
by keeping the same integrated spectrum (i.e. $T_\mathrm{eff}$). 
The difference between non-equilibrium and equilibrium chemistry 
is more important for GJ~504~b; around 10~$\mu$m we observe a 
difference of 50~$\%$ between equilibrium and non-equilibrium 
cases, but the normalised spectra of the non-equilibrium cases 
don't show a lot of variation (Fig.~\ref{GJ504bEmissNeq}). It 
seems to be easier to determine the $K_\mathrm{zz}$ with the 
VHS~1256--1257~b spectrum (spectra around 10 $\mu$m exhibit more 
variation in flux and the depth of NH$_3$ line is more dependent 
on the considered $K_\mathrm{zz}$, Fig.~\ref{VHSbEmissNeq}). 

In a low $T_\mathrm{eff}$ case (Fig.~\ref{GJ504bEmissNeq}), 
non-equilibrium chemistry has an important global effect weakly 
dependent on the eddy coefficient (the non--equilibrium effect is 
something like saturated). At higher $T_\mathrm{eff}$ 
(Fig.~\ref{VHSbEmissNeq}) non-equilibrium chemistry has less 
effect but deep and hot molecules are effectively transported, 
proportionally to the $K_\mathrm{zz}$, to the top of the 
atmosphere. For GJ~504~b (Fig.~\ref{GJ504bEmissNeq}) around 
10~$\mu$m the three spectra at non-equilibrium just seem to be 
shifted without any strong shape modification. In VHS~1256--1257~b 
(Fig.~\ref{VHSbEmissNeq}) the shape of the NH$_3$ features varies 
significantly with the $K_\mathrm{zz}$, with a significant effect 
of PH$_3$ on the low-wavelength wing.\\

PH$_3$ is the only molecule included in \emph{Exo-REM} and not in 
\emph{V12} with a profile impacted by non-equilibrium 
chemistry (it is not the case for the alkalis). In this 
paragraph we study the effect of PH$_3$ on the spectra.

\begin{figure*}[ht]
\begin{center}
\includegraphics[width=1\textwidth]{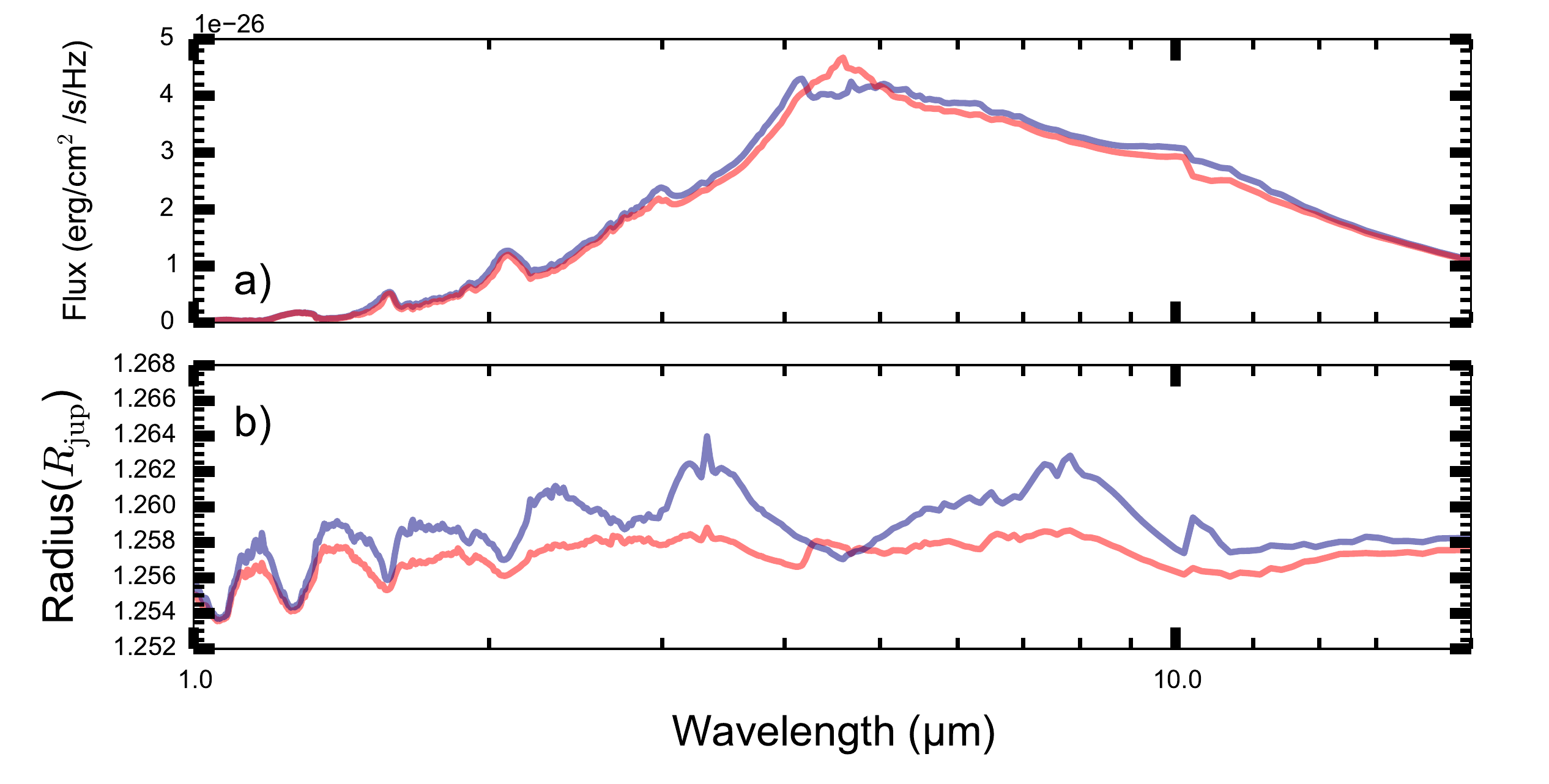}
\caption{{Emission (up) and transmission (bottom) spectra of the test exoplanet with a 
$T_\mathrm{eff}$=1000~K, computed by \emph{petitCODE} using equilibrium chemistry
(blue) or non equilibrium with K$_{zz}$=10$^{11}$~cm$^2$s$^{-1}$ (red) using \emph{V12}.
\label{transitneq}}}
\end{center}
\end{figure*}

\begin{figure}[ht]
\begin{center}
\includegraphics[width=1\columnwidth]{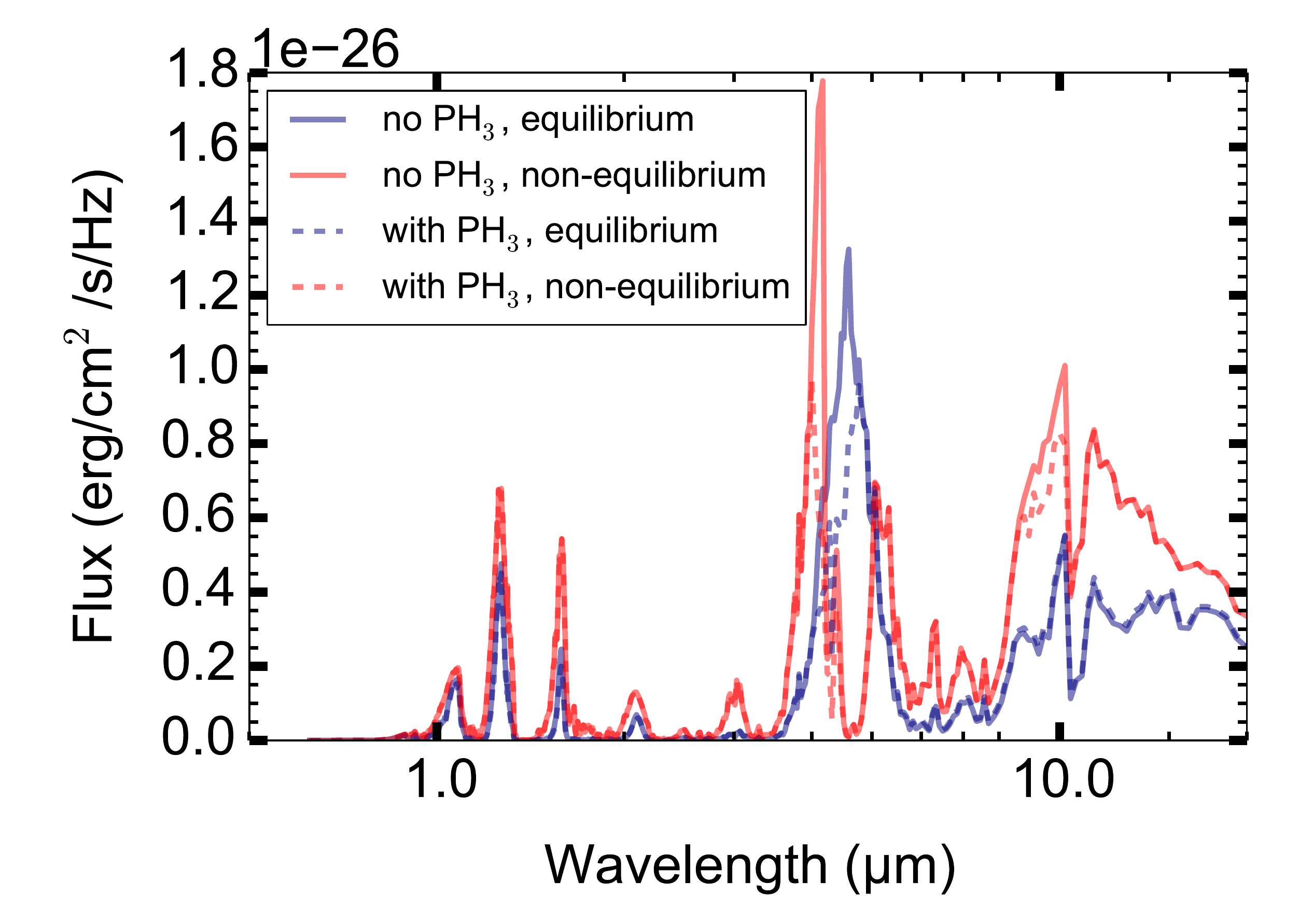}
\caption{{Spectra of GJ~504~b, computed by \emph{Exo-REM} with and without PH$_3$, assuming thermochemical equilibrium or in non-equilibrium conditions ($K_\mathrm{zz}$=10$^{11}$cm$^2$s$^{-1}$) with the same PT profile (case with PH$_3$ at equilibrium) \label{ph3effect}}}
\end{center}
\end{figure}

We compute first the full model of GJ~504~b with \emph{Exo-REM}, 
at equilibrium, with PH$_3$. Then we use the obtained PT profile 
in this first case in \emph{Exo-REM} to generate spectra at 
equilibrium or with K$_{zz}$=10$^{11}$cm$^2$/s, with or without 
PH$_3$. Figure~\ref{ph3effect} shows all the corresponding 
spectra. We see the effect of phosphine around 4.5 and 9~$\mu$m. 
At thermochemical equilibrium  the right wing of the PH$_3$ 
feature is visible at 4.5~$\mu$m but the signature of PH$_3$ at 
9~$\mu$m is completely hidden by the ammonia absorption. In the 
case of non--equilibrium chemistry both PH$_3$ features are 
visible, especially the left wing around 4.5~$\mu$m, but the red 
wing is hidden by CO and CO$_2$.

\section{Influence of the differing model assumptions on the analysis of JWST observations \label{detectability}}
In this section we quantify differences 
coming from the previously identified model differences and 
compare them with the expected direct imaging uncertainties of 
JWST observations. For this analysis we use the signal-to-noise 
ratio (S/N) that we previously estimated for the NIRSpec and MIRI 
instrument modes, described for Fig.~\ref{EmissTest} in 
Sec.~\ref{resultscomp} for VHS~1256--1257~b.

We compute the $\chi^2$ in the wavelength range where 
we observe an effect between the average spectrum (between all the 
approaches) and the other spectra taken separately. We perform, in 
parallel, a Kolmogorov–Smirnov test (KStest) as done in 
\cite{Molli_re_2017}. If the $\chi^2$ or the $p_\mathrm{KStest}$ 
is around 1, then the differences between the approaches 
correspond to the uncertainties of the JWST. If, on the other 
hand,  $\chi^2>>1$ or $p_\mathrm{KStest} << 1$, then the 
differences between the approaches are larger than the JWST 
uncertainties.\\

First, for the different alkali line profiles, 
we use the spectra plotted in 
Figs.~\ref{fig:alkali1}~and~\ref{fig:alkali4},
restrained to the wavelength range between 0.6 and 0.9 $\mu$m
(i.e. in the part of the NIRSpec range that is mostly impacted by 
alkalies).

We find that, in the case of directly imaged exoplanet, 
the average $\chi^2$, as defined as above, is always larger than 
20 and ($p_\mathrm{KStest}$ is lower than $10^{-30}$).

Therefore, the precision expected with JWST allows us to 
hope to be able to learn more about the profile of the alkalies 
even if this wavelength range is also impacted by the clouds (we don't 
study this effect in the paper).\\

We focus then on the effect of the selection of a given 
molecular far wing lineshape, using the spectra plotted in  
Fig.~\ref{fig:self_cons_emis_irrad}. We combine here wavelength 
ranges from NIRSpec and MIRI: 0.6-13.9 $\mu$m (the same as for 
VHS~1256--1257~b previously used in this paper).

For a directly imaged planet, we find that using or not a 
far wing cut-off yields a significant difference (average 
$\chi^2 > 20$ and $p_\mathrm{KStest} < 10^{-9}$).

The effect is more important in spectral windows outside 
the main molecular features, acting like a continuum, and may 
induce uncertainties in the retrievals of some atmospheric 
parameters, such as clouds parameters, larger than those due to 
the uncertainties on the data themselves.\\

For directly imaged planets, differences between 
equilibrium and out of equilibrium compositions 
(K$_{zz}$=10$^{11}$cm$^2$/s) are largely higher than the 
observational uncertainties (the average $\chi^2$ is over 15 and 
$p_\mathrm{KStest} < 10^{-4}$). It should be possible to determine 
what is the chemical regime of the observed atmosphere.\\ 

Finally we focus on PH$_3$, in two wavelength ranges, 4-6 
(NIRSpec) and 9-10 (MIRI) $\mu$m using Fig.~\ref{ph3effect}. This 
molecule has a strong effect on the observable spectra and it 
should be easy to discriminate whether the atmosphere contains 
PH$_3$ or not (with a minimal average $\chi^2$ of 40 and  
$p_\mathrm{KStest} < 10^{-10}$).

\section{Conclusions and perspectives \label{discuss}}
The forthcoming JWST will open a new era to characterize exoplanet
atmospheres. The observed spectra with the high signal-to-noise 
($>$~100 for exoplanets detected in direct imaging such as 
VHS~1256-1257~b) expected with the JWST instruments will require 
precise and well tested models to interpret the results. We need 
to be able to understand differences between models. To do that we 
define in this paper a benchmark protocol.

We compared and updated three radiative-convective 
equilibrium models (\emph{ATMO}, \emph{Exo-REM} and 
\emph{petitCODE}) and compared two chemical models (\emph{V12} and 
\emph{Exo-REM}). We showed that the analysis of JWST 
observations will be sensitive to differences in the modelling of 
some key parameters introduced in the various models. This fact 
should be kept in mind when drawing conclusions.\\

Many iterations were needed to achieve similar results in our 
benchmark conditions. We have spotted the importance of PH$_3$ in 
atmospheres, and also the crucial need of complete molecular 
linelists. Exomol is often considered as the most 
up-to-date database currently available, and is indeed better 
adapted to hot atmospheres than Hitran (built for Earth-like 
temperatures). However, the current version of Exomol is not the 
ultimate linelist database. Improvements are still going on, for 
example \cite{Rey2017} for CH$_4$. During 
the definition of the benchmark protocol, we spotted that 
chemistry model can also imply important differences, if one takes 
into account ionic species at hot temperature, isotopes, etc.\\

The first major difference spotted early in the process was the 
different treatment applied for the far wing lineshape of alkali 
lines. The definition of the alkali treatment is crucial. Using 
one approach or the other can shift the temperature by more than 
100~K and change the flux in spectral region where we also expect 
clouds effects \citep[reddening of the spectrum observed for example in brown dwarf observations][]{1996A&A...305L...1T}. 
However, comparison to observations does not allow us to select 
the best approach.\\

We investigated also a second effect. We found that applying a 
sub-Lorentzian lineshape in the far wings of molecular species can 
change the resulting spectra both in transmission and emission by 
up to $\sim 20$~\%, but usually less. This is comparable to what 
has been found by \cite{Grimm_2015}. In the self-consistent 
temperature structures obtained for irradiated planets, an 
increased opacity (due to neglecting the cut-off) leads to an 
inverse green house effect in the atmosphere, because less stellar 
radiation is absorbed in the deep atmosphere if full Voigt 
profiles are used. Uncertainties in the far wing lineshape 
is also a limitation in analysing thermal emission spectra from 
Solar System planetary atmospheres. This is the case for 
relatively transparent spectral windows, such as the methane 
windows in Titan between 1.3 and 5~$\mu$m 
\citep[e.g.][]{Hirtzig2013} or near-infrared windows observed on 
Venus' night side between 1 and 2.3~$\mu$m 
\citep[e.g.][]{Fedorova2015}. Sub-lorentzian lineshapes beyond a 
few wavenumbers from line core are clearly needed to reproduce the 
observed spectra but, precise constraints on the profiles cannot 
be retrieved independently of other atmosphere or surface 
parameters. For giant exoplanets, while sub-Lorentzian far wing 
profiles are also likely to occur, no prescription can be made 
today in the lack of laboratory measurements of H$_2$-broadened 
line far wings at high temperatures.

To compare models carefully we need to use the same alkali 
treatment and far wing lineshape for other absorbers.\\

We have also compared various chemical approaches. While we see  
clear differences between non-equilibrium and equilibrium 
abundances, we did not spot significant differences in the spectra 
based on the different implementations used for the two processes. 
The major differences occur at high altitudes: cold-trap and 
molecular scattering, without strong impact on the spectrum..\\

Among all the models parameters discussed and analysed in 
the paper, those having a major impact on the observable are:
\begin{itemize}
 \item Alkali wing lineshape
 \item PH$_3$ opacity
 \item sub-Lorentzian lineshape
 \item non-equilibrium chemistry
\end{itemize}
We showed that these four modelling parameters will have 
an impact on spectra with differences larger than the JWST 
uncertainties, especially for directly imaged exoplanets. However, 
in the case of alkali lineshape or molecular line far wing, the 
differences in the modelled spectra could be translate as an 
artificial off-set, without an easy way to identify it.\\

To improve the reliability of atmospheric models, 
particular efforts need to be undertaken to enhance experimental 
and theoretical work at high pressure and high temperature 
conditions in terms of :
\begin{itemize}
   \item wing lineshape, especially for alkalies
   \item linelist completeness
\end{itemize}
This work is urgently needed if we want to be able to 
interpret correctly the observations of the JWST.

We also identified some other effects coming from:
\begin{itemize}
 \item adding isotopes
 \item adding ions
 \item using up-to-date linelists
 \item definition of the reference radius for transmission 
 spectrum
 \item computation of the mean molecular weight
\end{itemize} 
And finally the differences, spotted in the abundances profiles, 
without any significant spectral effect found in the analysis are:
\begin{itemize}
 \item cold-trap
 \item molecular diffusion
\end{itemize}

Spectra and profiles generated for the benchmark (i.e. the data behind the 
Figs.~\ref{ColdGuillotSpectra}-\ref{targetsPT} of the 
Appendix~\ref{allfigures}) are available as supplementary material. 
We encourage the community to compare them to their models in these 
benchmark conditions and iterate with us to continue to improve our 
respective models and identify the existing differences.

\section*{Acknowledgements}
We thank the referee to her/his fast report and for the useful comments.
POL acknowledges support from the LabEx P2IO, the French ANR 
contract 05-BLAN-NT09-573739. OV acknowledges support from the KU 
Leuven IDO project IDO/10/2013, the FWO Postdoctoral Fellowship 
program, and the Centre National d'Etudes Spatiales (CNES).

\appendix

\section{Complete set of figures \label{allfigures}}
Following figures 
(Figs.~\ref{ColdGuillotSpectra}-\ref{abundVHSneq}) show all the 
radiative-convective equilibrium models in the benchmark condition 
(see Sects.~\ref{benchmarkcond}~and~\ref{benchmark}).

\begin{figure}[ht]
\begin{center}
\includegraphics[width=1\columnwidth]{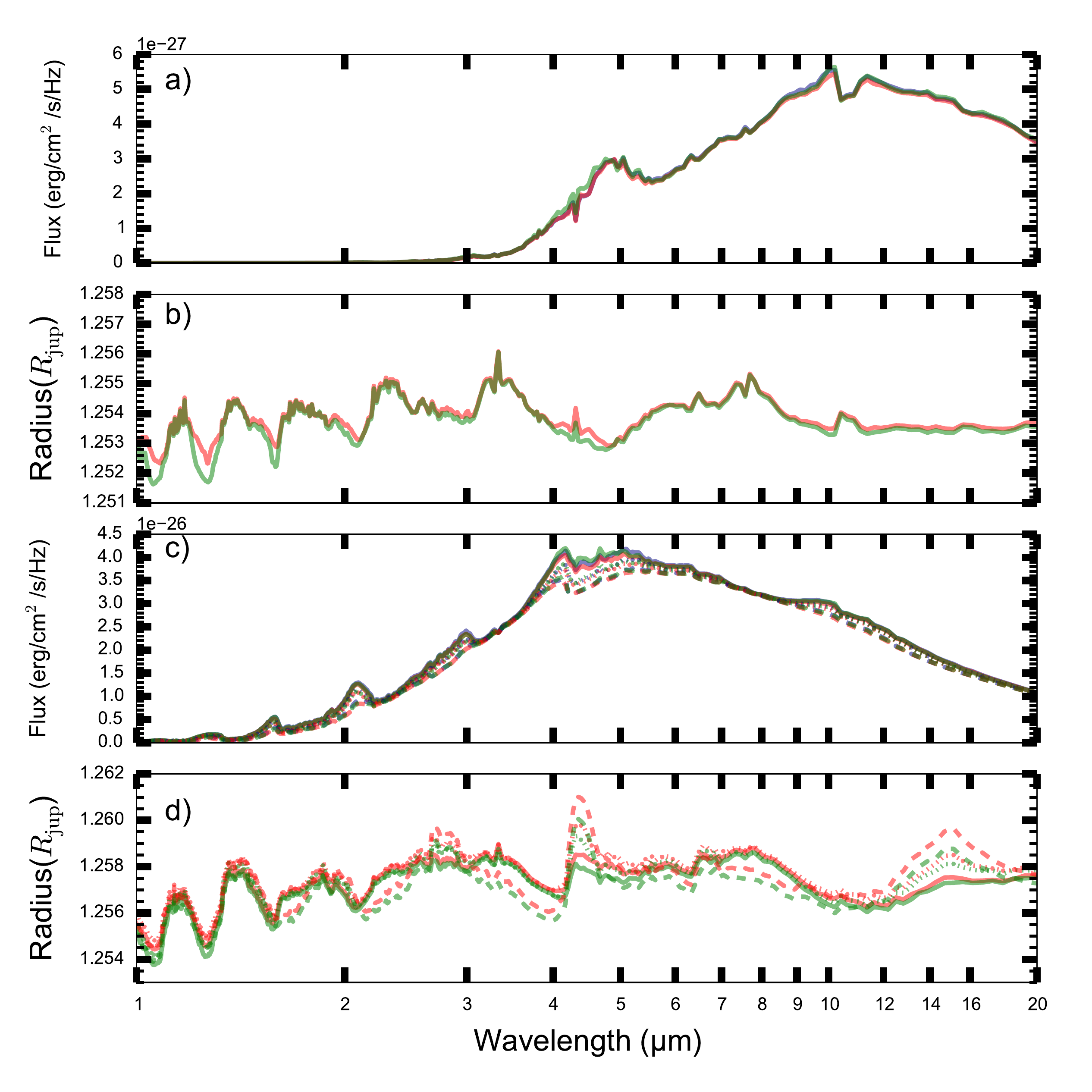}
\caption{{Emission and transmission spectra for the case T$_\mathrm{eff}$ = 500~K ( respectively a) and b) ) and the cases T$_\mathrm{eff}$ = 1000~K ( respectively c) and d) ) with variation of metallicity ( solid = solar, small dash = 3 $\times$ solar, large dash = 30 $\times$ solar ). Exo-REM is in blue, petitCODE in green, ATMO in red.\label{ColdGuillotSpectra}}}
\end{center}
\end{figure}

\begin{figure}[ht]
\begin{center}
\includegraphics[width=1\columnwidth]{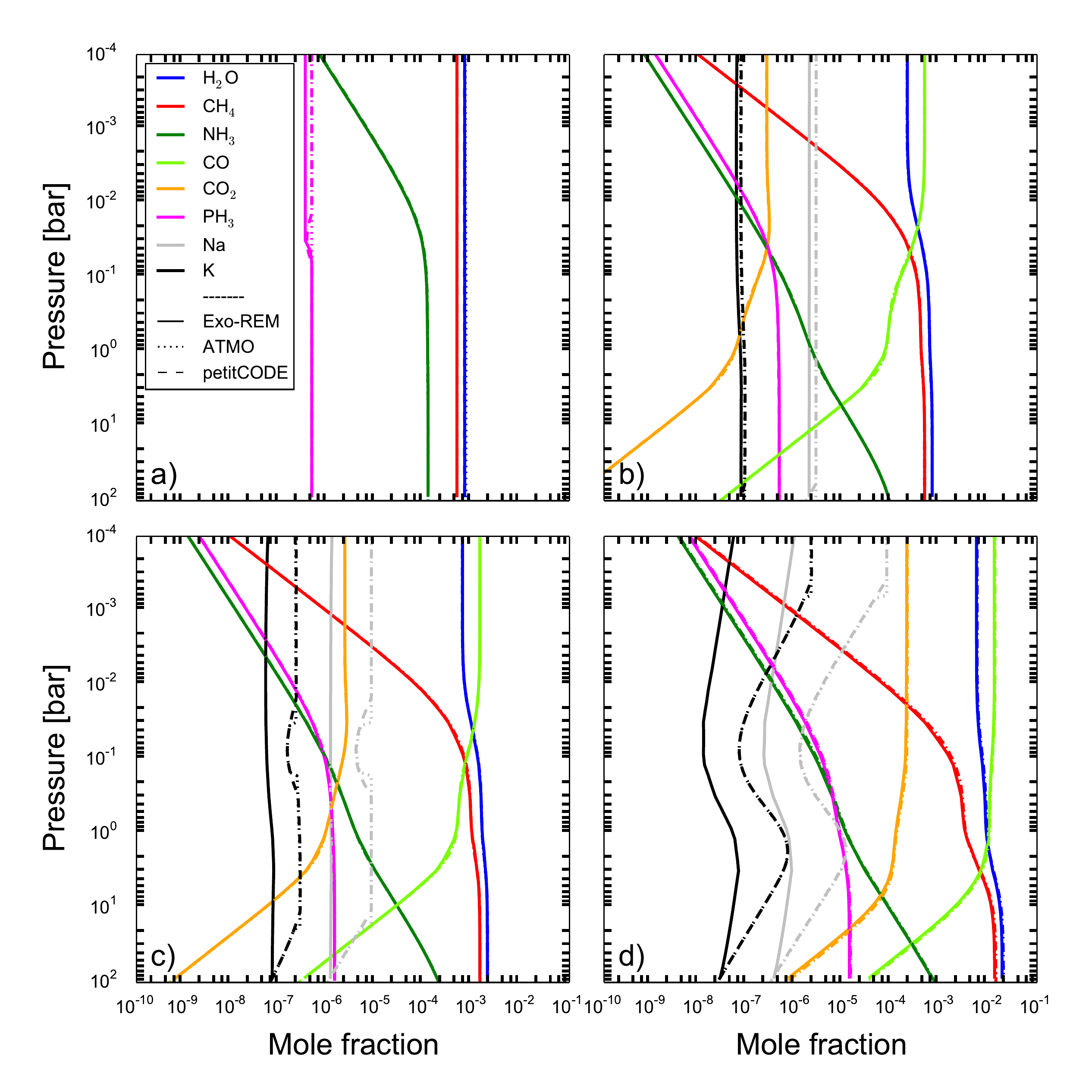}
\caption{{Abundance profiles for the case Teff = 500 K ( a) ) and the cases Teff = 1000 K with variation of metallicity ( b)solar, c) 3 $\times$ solar, d) 30 $\times$ solar).\label{ColdGuillotAbund}}}
\end{center}
\end{figure}

\begin{figure}[ht]
\begin{center}
\includegraphics[width=1\columnwidth]{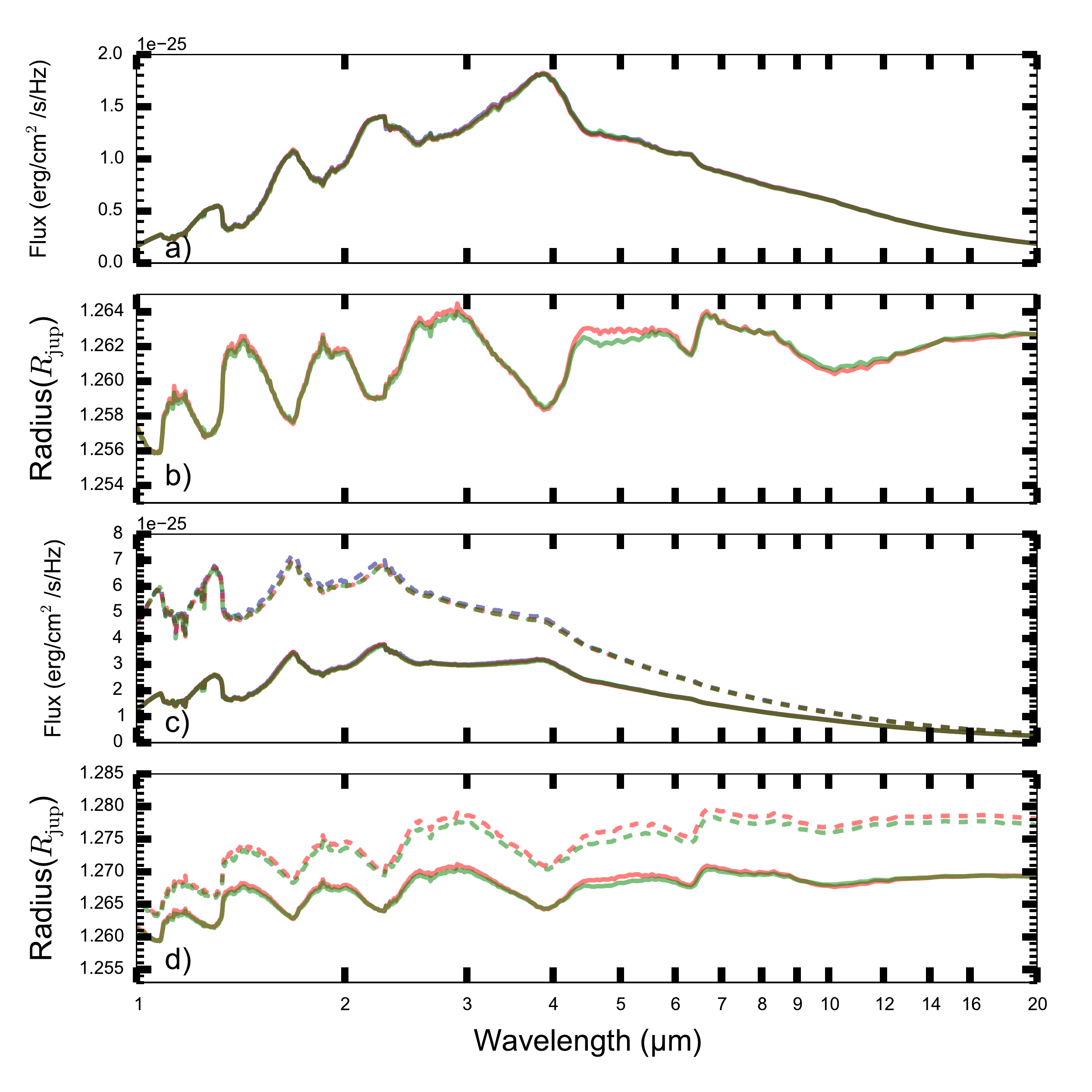}
\caption{{Emission and transmission spectra for the case T$_\mathrm{eff}$ = 1500~K ( respectively a) and b) ) and the cases T$_\mathrm{eff}$ = 2000~K (plain) and 2500~K (dash) ( respectively c) and d) ). Exo-REM is in blue, petitCODE in green, ATMO in red.\label{HotGuillotSpectra}}}
\end{center}
\end{figure}

\begin{figure}[ht]
\begin{center}
\includegraphics[width=1\columnwidth]{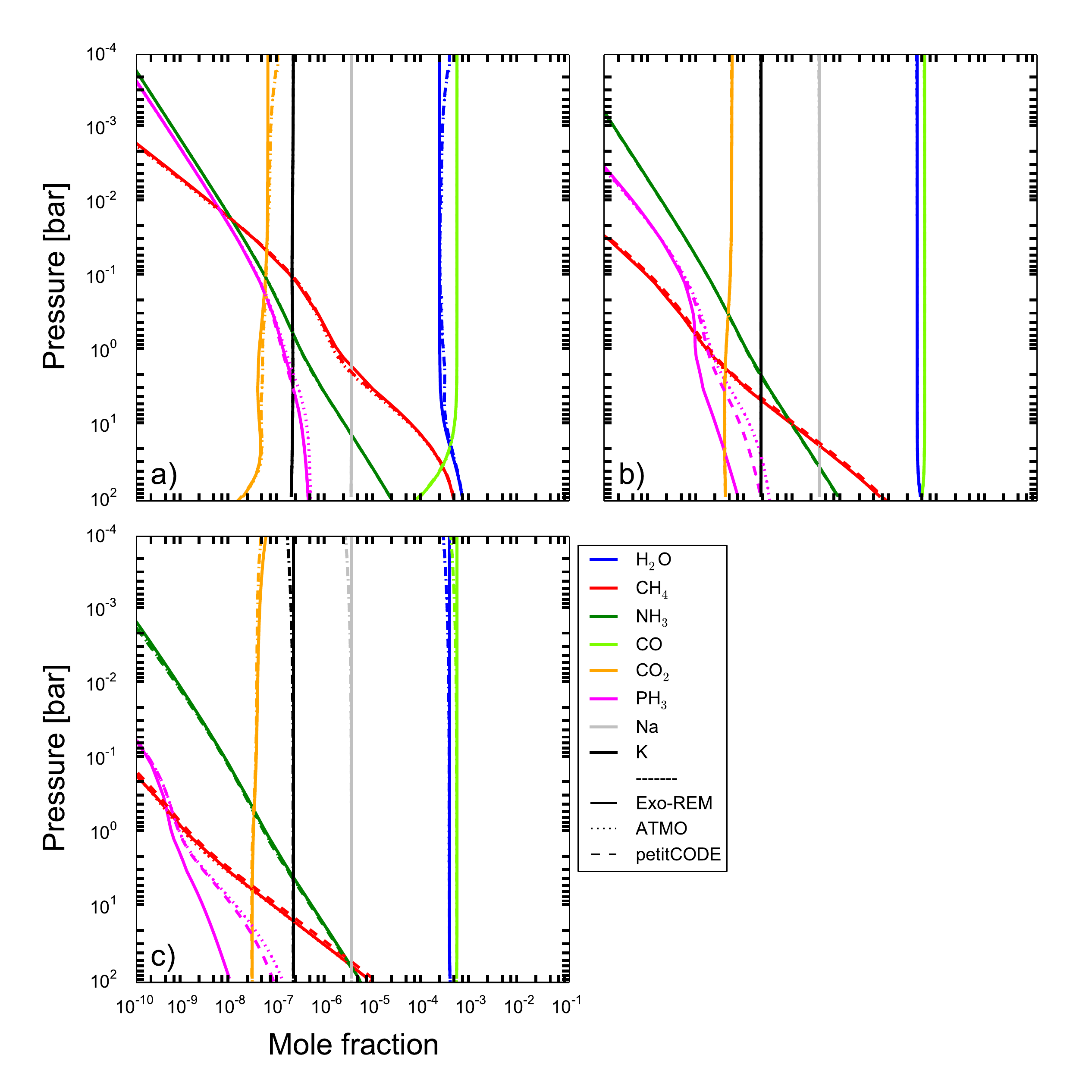}
\caption{{Abundance profiles for the cases T$_\mathrm{eff}$ = 1500~K ( a) ), 2000~K ( b) ) and 2500~K ( c) ).\label{HotGuillotAbund}}}
\end{center}
\end{figure}

\begin{figure}[ht]
\begin{center}
\includegraphics[trim = 0 20 0 0, clip,width=1\columnwidth]{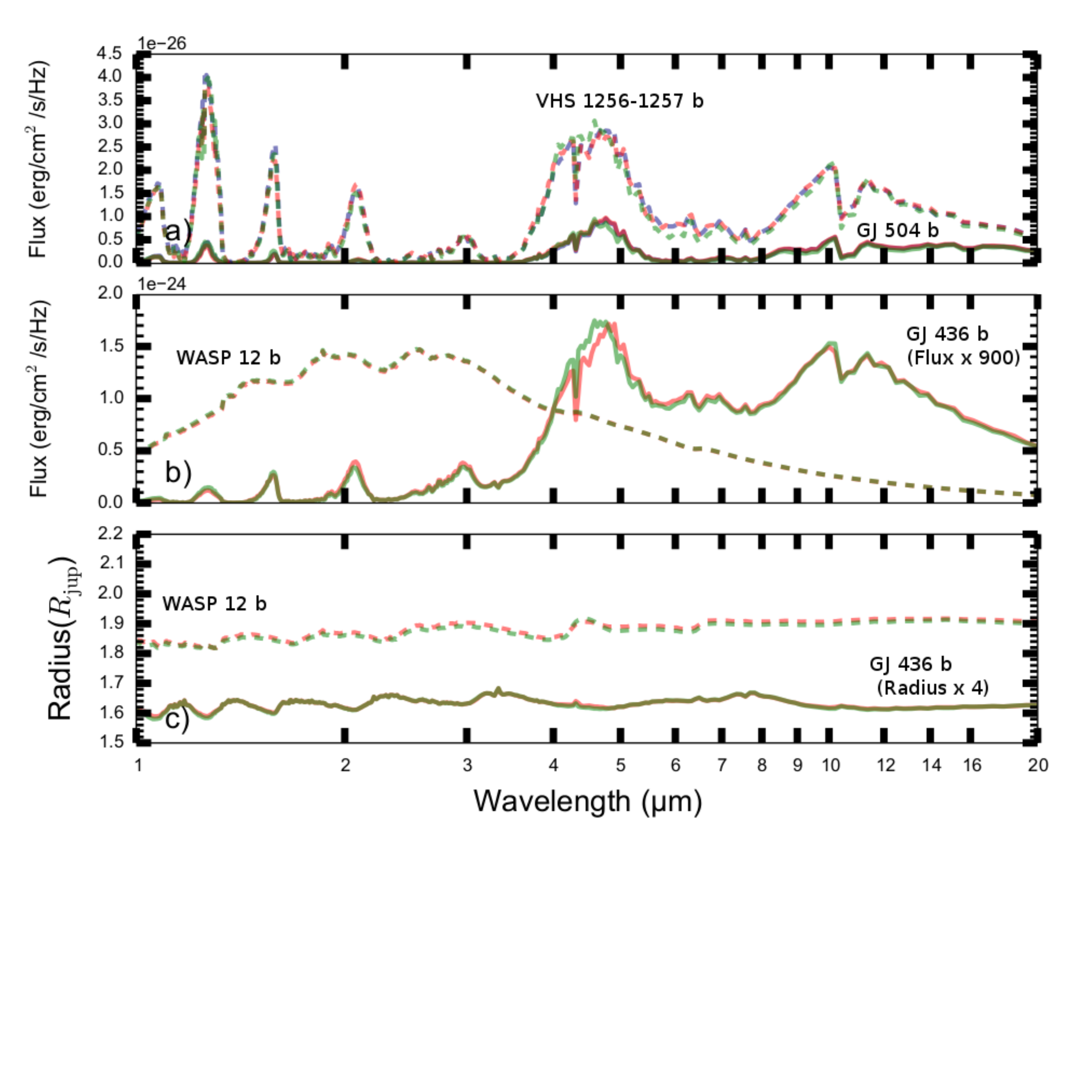}
\caption{{Emission spectra of the two directly imaged planets ( a) ). Emission ( b) ) and transmission ( c) ) spectra of the two transiting planets. For GJ~436~b the flux is multiplied by 900, and the radius by 4 to put the two cases in the same plot.\label{targetsSpec}}}
\end{center}
\end{figure}

\begin{figure}[ht]
\begin{center}
\includegraphics[width=1\columnwidth]{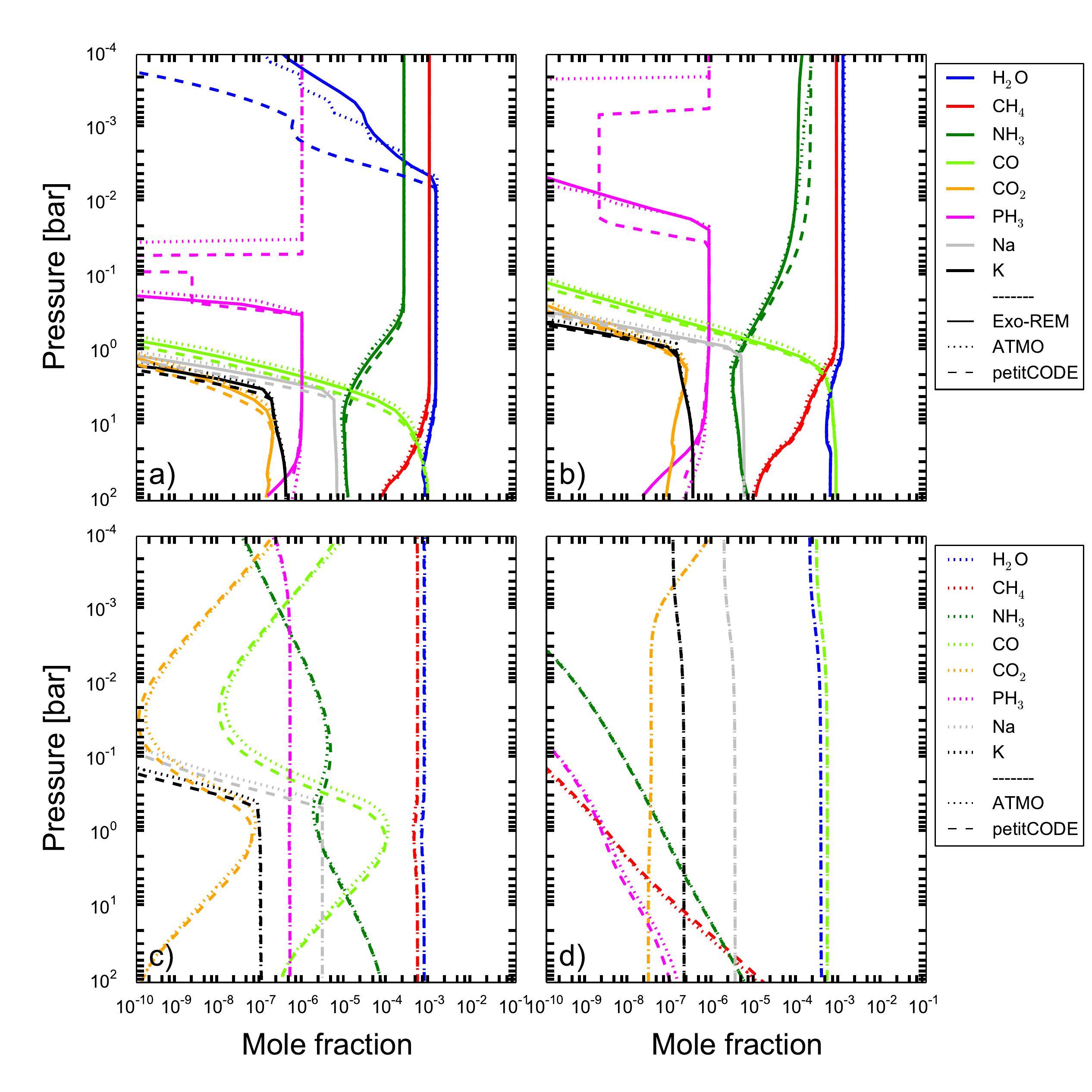}
\caption{{Abundance profiles of the potential targets: a) GJ~504~b, b) VHS~1256--1257~b, c) GJ~436~b and d) WASP~12~b.\label{targetsAbund}}}
\end{center}
\end{figure}

\begin{figure}[ht]
\begin{center}
\includegraphics[width=1\columnwidth]{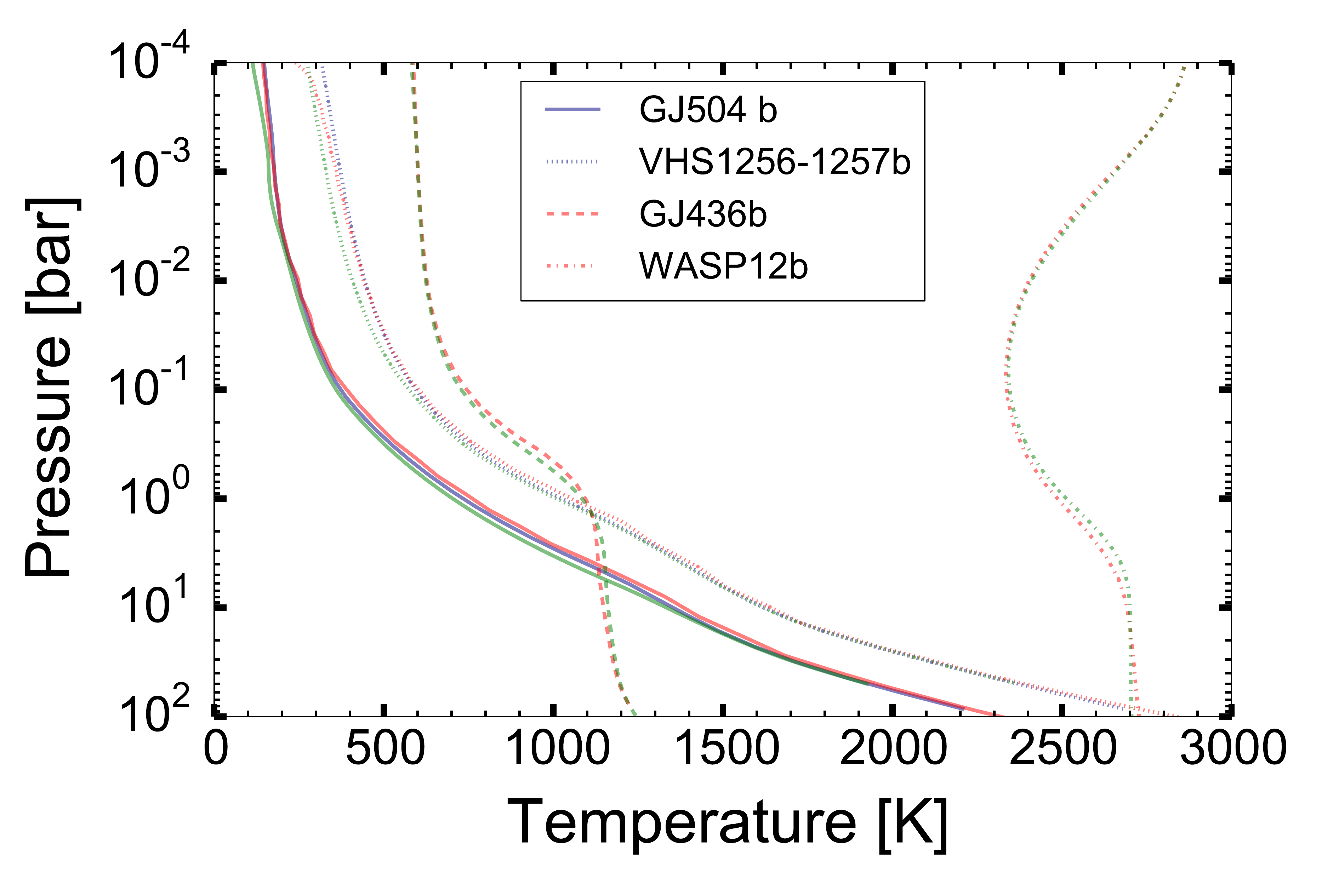}
\caption{{Temperature profiles computed by our three models (Exo-REM is in blue, petitCODE in green, ATMO in red) of the potential targets.\label{targetsPT}}}
\end{center}
\end{figure}

They complement figures shown in Sect.~\ref{nonequilibrium}

\begin{figure}[ht]
\begin{center}
\includegraphics[width=1\columnwidth]{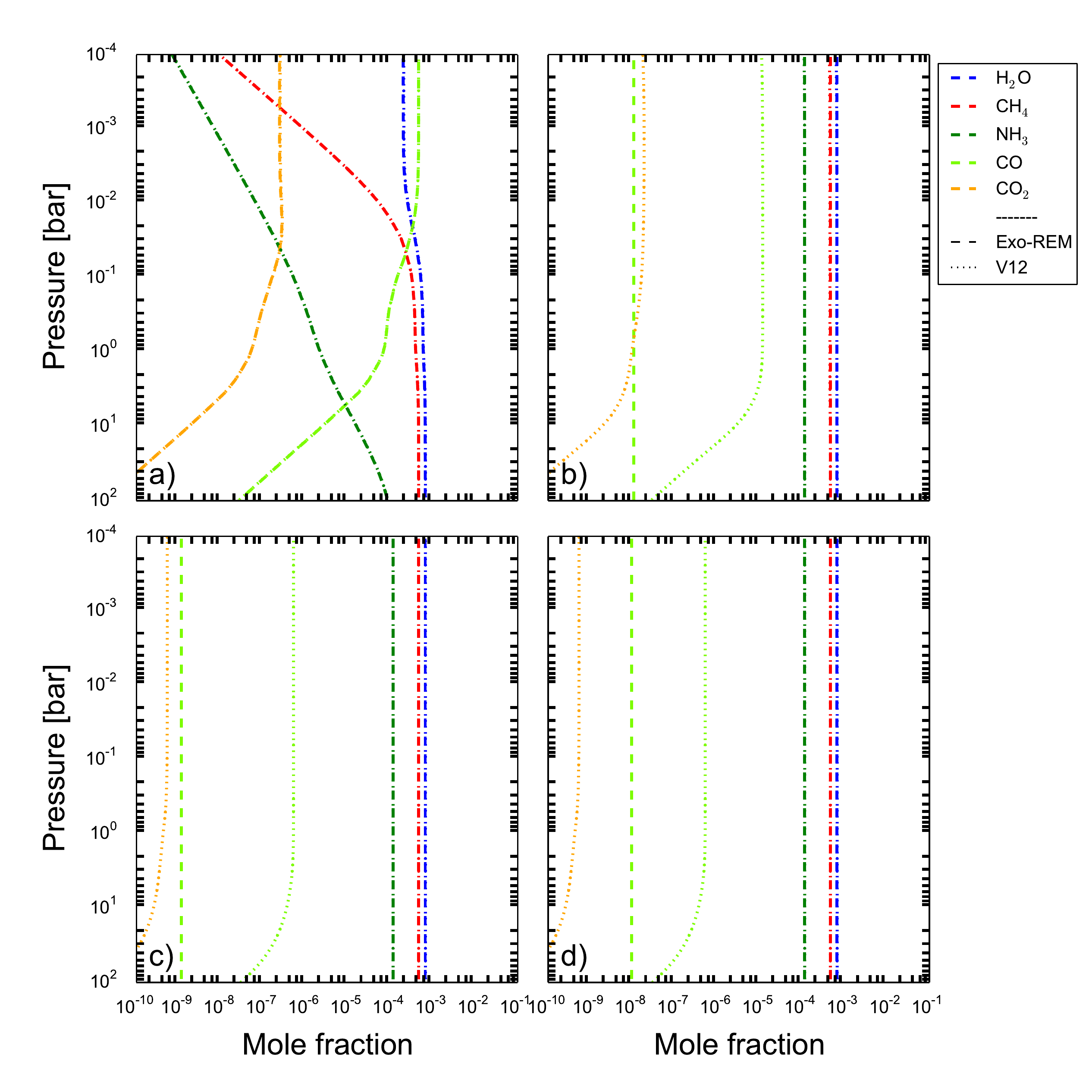}
\caption{{Abundance profiles for T$_\mathrm{eff}$ = 1000~K at equilibrium chemistry ( a) ) and for  K$_\mathrm{zz}$ = 10$^{7}$ ( b) ), 10$^{9}$ ( c) ) and 10$^{11}$ ( d) ) cm$^2$ s$^{-1}$\label{1000neqAbund}}}
\end{center}
\end{figure}

\begin{figure}[ht]
\begin{center}
\includegraphics[width=1\columnwidth]{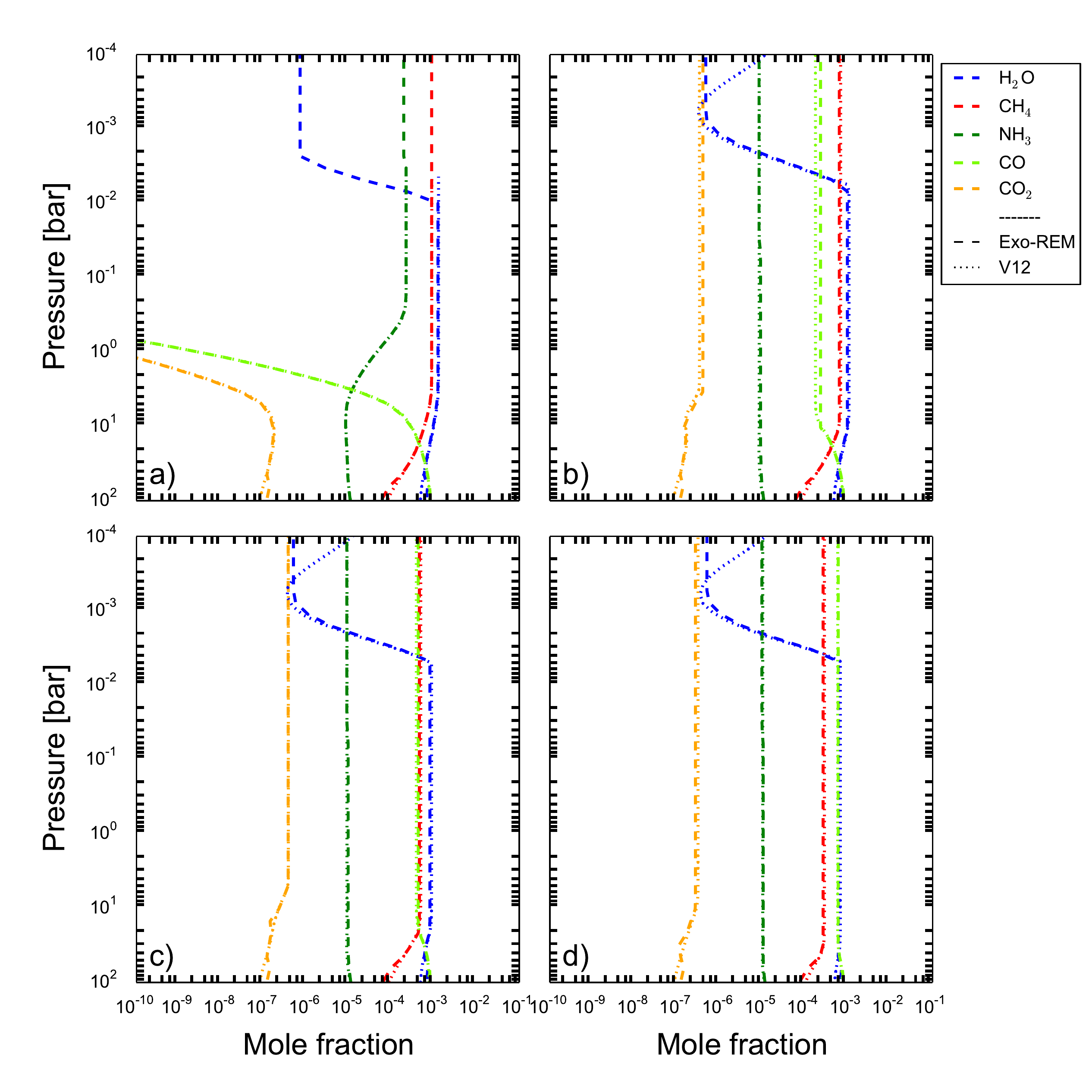}
\caption{{Abundance profiles for GJ~504~b at equilibrium chemistry ( a) ) and for  K$_\mathrm{zz}$ = 10$^{7}$ ( b) ), 10$^{9}$ ( c) ) and 10$^{11}$ ( d) ) cm$^2$ s$^{-1}$\label{GJ504neqAbund}}}
\end{center}
\end{figure}

\begin{figure}[ht]
\begin{center}
\includegraphics[width=1\columnwidth]{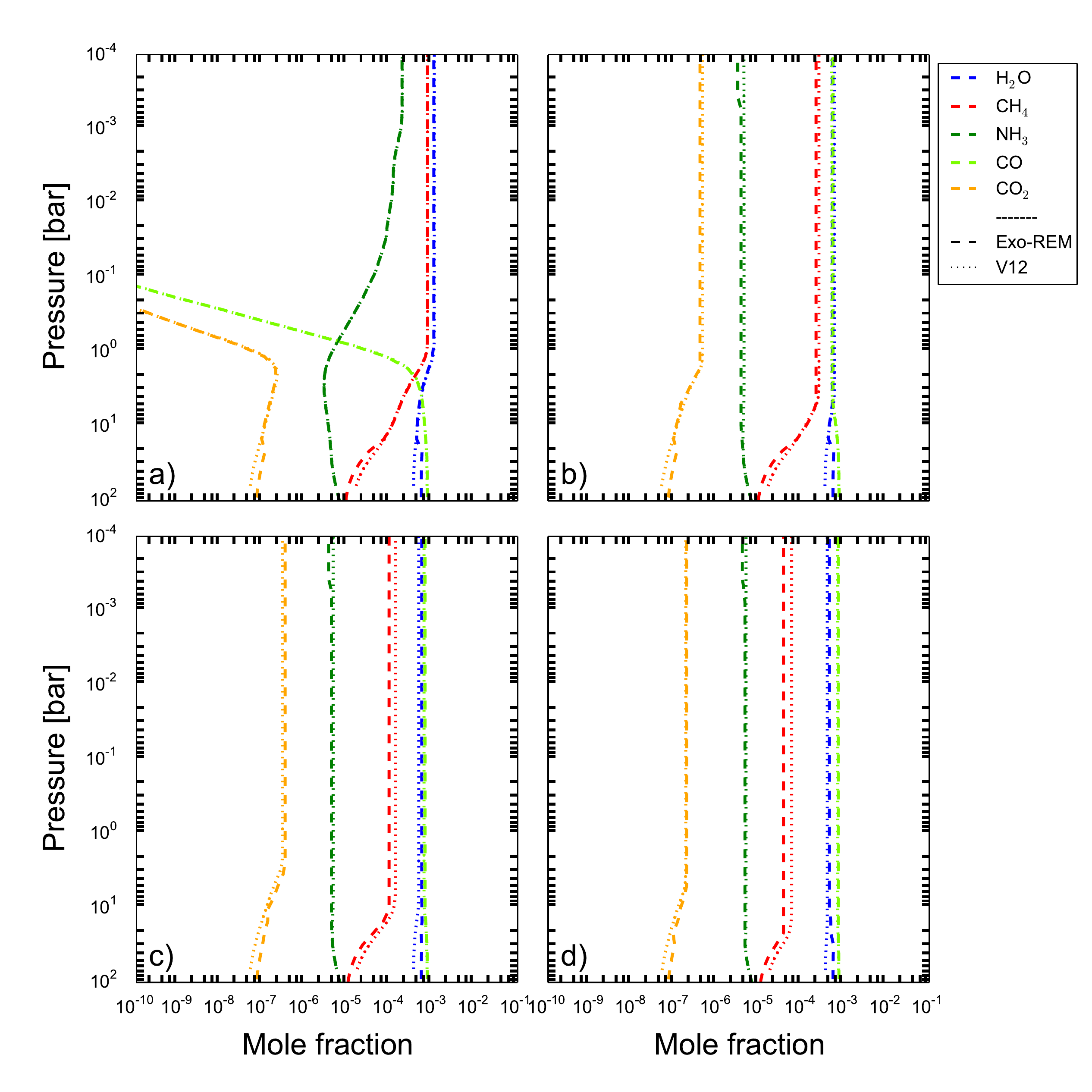}
\caption{{Abundance profiles for VHS~1256--1257~b at equilibrium chemistry ( a) ) and for  K$_\mathrm{zz}$ = 10$^{7}$ ( b) ), 10$^{9}$ ( c) ) and 10$^{11}$ ( d) ) cm$^2$ s$^{-1}$ \label{abundVHSneq}}}
\end{center}
\end{figure}

\section{Updates of \emph{Exo-REM} \label{upexo}}

\emph{Exo-REM} has been updated on several aspects since the version described in \citet{Baudino_2015}:

\begin{itemize}
\item Correlated k-coefficients of different molecules are now 
combined assuming no correlation between species in any spectral 
interval. This is done using the 
method named "reblocking of the joint k-distribution" by 
\citet{Lacis_1991} and more extensively described in 
\citet{Molli_re_2015} ("R1000" method in their Appendix B.2.1)

\item CO$_2$ and PH$_3$ have been added to the list of molecular 
absorbers, using linelists from \citet{Rothman_2013} and 
\citet{Sousa_Silva_2014} respectively. For CO$_2$ 
pressure-broadened halfwidths ($\gamma$), we used the 
air-broadened values multiplied by 1.34 \citep{Burch_1969} to 
obtain the broadening by H$_2$ and He. For PH$_3$, we used: for $J 
\leq 20, \gamma = 0.1009 - 0.00137 \times J$ if $J \neq K$ and 0.9 
time this result if J=K with a temperature exponent 
$n = 0.643 - 0.00912 \times J$; for $J > 20$, we fixed 
$\gamma = 0.073$ and $n = 0.46$. Relevant references are 
\citet{Levy_1993}, \citet{Salem_2004} and \citet{Levy_1994}.

\item Collision-Induced Absorption (CIA) of H$_2$-H$_2$ and 
H$_2$-He is now modeled following HITRAN \citep{Richard_2012}, 
complemented for H$_2$-H$_2$ beyond 10000 cm$^{-1}$ by 
\citet{Borysow_2001} and \citet{Borysow_2002} with a small 
rescaling to ensure continuity at 10000 cm$^{-1}$.
\end{itemize}

%
%

\bibliography{converted_to_latex.bib}



\end{document}